\documentclass[rmp]{revtex4}

\usepackage[utf8]{inputenc}

\usepackage{amsmath}
\usepackage{amssymb}
\usepackage{graphicx}
\usepackage{color}
\usepackage{algorithm}
\usepackage[noend]{algpseudocode}
\usepackage{listings}
\usepackage[hidelinks]{hyperref}

\def\mean#1{\left\langle #1 \right\rangle}
\def\vec#1{\mathbf{#1}}

\def\kk{\vec{k}}

\def\rr{\vec{r}}
\def\ss{\vec{s}}
\def\nn{\vec{n}}

\def\tauint{\tau_\text{int}}

\def\CC{\hat C^{(c)}}
\def\sampn#1#2{#1^{(#2)}}

\begin{document}

\title{Correlation functions as a tool to study collective behaviour
  phenomena in biological systems}

\author{Tom\'as S. Grigera}

\affiliation{Instituto de F\'\i{}sica de L\'\i{}quidos y Sistemas Biol\'ogicos (IFLYSIB), 
CONICET y Universidad Nacional de La Plata, Calle 59 no.~789, B1900BTE
La Plata, Argentina}
\affiliation{ CCT CONICET La Plata, Consejo Nacional de Investigaciones
  Cient\'\i{}ficas y T\'ecnicas, Argentina}
\affiliation{Departamento de F\'isica, Facultad de Ciencias Exactas,
  Universidad Nacional de La Plata, Argentina}

\date{July 2, 2021}

\begin{abstract}

  Much of interesting complex biological behaviour arises from
  collective properties.  Important information about collective
  behaviour lies in the time and space structure of fluctuations
  around average properties, and two-point correlation functions are a
  fundamental tool to study these fluctuations.  We give a
  self-contained presentation of definitions and techniques for
  computation of correlation functions aimed at providing students and
  researchers outside the field of statistical physics a practical
  guide to calculating correlation functions from experimental and
  simulation data.  We discuss some properties of correlations in
  critical systems, and the effect of finite system size, which is
  particularly relevant for most biological experimental systems.
  Finally we apply these to the case of the dynamical transition in a
  simple neuronal model,

\end{abstract}

\maketitle


\section{Introduction}

Biological systems behave in complex ways.  This is particularly
evident at the level of whole organisms: animals can adapt to a
variety of situations, and this capacity is related to the large
number of states (spatiotemporal patterns) that their brains can
explore.  These states are realised on some physical substrate
(neurons in the case of the brain), but it is quite plausible that the
same kind of behaviour can be produced by another system, composed of
units that interact among themselves in a similar way, but
microscopically of a very different nature.  For example, robot birds
could interact to form flocks that look macroscopically like starling
murmurations, or a collection of neurons simulated in software, or
built with electronics, could produce the same kind of patterns and
behaviour as a brain.  The point is that many of the interesting
phenomenology of biological systems stems from collective behaviour
\cite{kaneko_life_2006}, and that much of it may arise independently
of the details of the biological units making up the system.  A first
description of collective properties starts of course with computation
of average global quantities.  But in complex systems there is much
information on how fluctuations around those averages are structured
in space and time.  Correlation functions, the subject of this
article, can capture information about these fluctuations, and thus
constitute one of the basic statistical physics tools to describe
collective properties.

To be sure, there is more to collective behaviour than what is
captured by the two-point correlation functions we discuss here.  But
correlation functions should be in the toolbox of any researcher
attempting to understand the behaviour of complex biological systems,
and it is probably fair to say that, although they have been used and
studied for a long time in statistical physics, they have not been
exploited enough in biology.

Critical systems certainly deserve mention among systems exhibiting
nontrivial collective behaviour.  At a critical point (i.e.\ at the
border between two static or dynamic phases) of a continuous
transition, long-range correlated fluctuations dominate the
large-scale phenomenology, so that many critical systems look alike at
large scales: this is called universality.  Criticality is rare, in
the sense that systems are critical only in a negligible region of the
space of their control parameters.  However, criticality appears to
play an important role in biological systems \citep{bak1996, Mora2011,
  munoz2018} and in particular the brain
\citep{chialvo_emergent_2010}, with different signs of critical
behaviour having been found in systems as diverse as proteins
\citep{mora2010, tang_critical_2017}, membranes \citep{veatch2007,
  honerkamp-smith_introduction_2009}, bacterial colonies
\citep{dombrowski2004, zhang2010}, social amoebas
\citep{de_palo_critical-like_2017}, networks of neurons
\citep{beggs2003, schneidman2006}, brain activity
\citep{fraiman_what_2012, haimovici_brain_2013}, midge swarms
\citep{Attanasi2014, cavagna_dynamic_2017}, starling flocks
\citep{Cavagna2010} and sheep herds \citep{ginelli_intermittent_2015}.
Since the odds that the system's parameters are tuned to critical just
by chance are vanishing, the question arises of why critical systems
appear to be so common in biology.  An answer may be that criticality
is the simple path to complexity \citep{chialvo2018} and thus the
reason why a functioning brain, for instance, needs to be ``at the
boundary between being nearly dead and being fully epileptic''
\citep{Mora2011}.  Two-point correlation functions are an important
tool in the exploration of critical systems.  Again, they are not the
whole story, but a necessary instrument nevertheless.

The purpose of this contribution is twofold: for one, it aims to
present in a brief but self-contained form, the definitions of several
two-point correlation functions together with a discussion on how to
compute them from experimental or simulation data.  And second, to
argue that a simultaneous study of space and time correlations can
yield useful information, and perhaps an alternative view, of the
dynamic behaviour of neural networks.  We will illustrate this on a
simple neural model.

In the following, we present the theoretical definition of two-point
correlation functions (\S\ref{sec:definitions}), followed by a
discussion on their computation from experimental data
(\S\ref{sec:comp-corr-funct}). Then we discuss the definition and
computation of the correlation length and time
(\S\ref{sec:relax-time-corr}) and some properties of the correlation
functions of near-critical systems (\S\ref{sec:corr-funct-crit}).
Finally we present some results on space and time correlations on a
neural model (\S\ref{sec:dynam-scal-neur}).


\section{Definitions}
\label{sec:definitions}

\subsection{Correlation and covariance}
\label{sec:covar-corr}

Let $x$ and $y$ be two random variables and $p(x,y)$ their joint
probability density.  We use $\langle\ldots\rangle$ to represent the
appropriate averages, e.g.\ the mean of $x$ is
$\mean{x}=\int x \, p(x) dx$ (probability distribution
of $x$ can be obtained from the joint probability,
$p(x)=\int\!\!dy\,p(x,y)$, and in this context is called marginal
probability) and its variance is
$\text{Var}_x=\bigl\langle (x-\mean{x})^2\bigr\rangle$.  We define
\begin{align}
  C_{xy} &= \langle x y \rangle = \int x y \, p(x,y)
           dxdy,   \label{eq:corr} \\
  \text{Cov}_{x,y} &= \Bigl\langle \bigl( x-\mean{x}\bigr)
  \bigl(y-\mean{y}\bigr) \Bigr\rangle =
  \langle x y \rangle - \mean{x}\mean{y}.
  \label{eq:covariance}
\end{align}
We call $C_{xy}$ the \emph{correlation} and $\text{Cov}_{x,y}$ the
\emph{covariance} of $x$ and $y$.  The covariance is bounded by the
product of the standard deviations \citep[\S 2.12]{Priestley1981},
\begin{equation}
  \label{eq:cov-ineq}
  \text{Cov}^2_{x,y} \le \text{Var}_x \text{Var}_y.
\end{equation}
and is related to the variance of the sum:
\begin{equation}
  \text{Var}_{x+y} = \text{Var}_x + \text{Var}_y + 2\text{Cov}_{x,y}.
  \label{eq:varsum}
\end{equation}

The \emph{Pearson correlation coefficient} is defined as
\begin{equation}
  r_{x,y} = \frac{\text{Cov}_{x,y}}{\sqrt{\text{Var}_x\text{Var}_y}},
\end{equation}
and, as a consequence of~\eqref{eq:cov-ineq}, is bounded:
$-1\le r_{x,y}\le1$.  It can be shown that the equality holds only
when the relationship between $x$ and $y$ is linear \citep[\S
2.12]{Priestley1981}.

The variables are said to be \emph{uncorrelated} if their covariance
is null:
\begin{equation}
  \label{eq:uncorrelated-def}
  \text{Cov}_{x,y}=0 \quad \Longleftrightarrow \quad
  \mean{xy} = \mean{x}\mean{y} \qquad \text{(uncorrelated)}.
\end{equation}
Absence of correlation is weaker than \emph{independence:}
independence means that $p(x,y)=p(x)p(y)$ (and clearly implies absence
of correlation).  For uncorrelated variables it holds, because
of~\eqref{eq:varsum}, that the variance of the sum is the sum of the
variances, but $\text{Cov}_{x,y}=0$ is equivalent to independence only
when the joint probability $p(x,y)$ is Gaussian.  The covariance, or
the correlation coefficient, are said to measure the degree of
\emph{linear association} between $x$ and $y$, because it is possible
to build a nonlinear dependence of $x$ on $y$ that yields zero
covariance \citep[see Ch.~2 of][]{Priestley1981}.


\subsection{Fluctuating quantities as stochastic processes}
\label{sec:stochastic}

Consider now an experimentally observable quantity, $a$, that provides
some useful information on a property of a system of interest.  We
assume here for simplicity that $a$ is a scalar, but the present
considerations can be rather easily generalized to vector quantities.
$a$ can represent the magnetization of a material, the number of
bacteria in a certain region, neural activity (e.g.\ as measured by
fMRI), etc.  We assume that $a$ is a local quantity, i.e.\ that its
value is defined at a particular position in space and time, so that
we deal with a function $a(\rr,t)$ (for the case where only space or
time dependence is available see \S\ref{sec:glob-time-indep}).

We are interested in cases where $a$ is \emph{noisy,} i.e.\ subject to
random fluctuations that arise because of our incomplete knowledge of
the variables affecting the system's evolution, or because of our
inability control them with enough precision (e.g.\ we do not know all
variables that can affect the variation of the price of a stock market
asset, we do not know all the interactions in a magnetic system, we
cannot control all the microscopic degrees of freedom of a thermal
bath).  We wish to compare the values of $a$ measured at different
positions and times, but a statement like ``$a(\rr_1,t_1)$ is larger
than $a(\rr_2,t_2)$'' is useless in practice, because even if it is
meaningful for a particular realization of the measurement process,
the noisy character of the observable means that a different
repetition of the experiment, under the same conditions, will yield a
different function $a(\rr,t)$.  Actually repeating the experiment may
or may not be feasible depending on the case, but we assume that we
know enough about the system to be able to assert that a hypothetical
repetition of the experiment would not exactly reproduce the original
$a(\rr,t)$.  For clarity, it may be easier to imagine that several
copies of the system are made and let evolve in parallel under the
same conditions, each copy then producing a signal slightly different
from that of the other copies.  The quantity $a(\rr,t)$ is then a
random variable, and to compare it at different values of its argument
we will use \emph{correlation functions,} and this section is devoted
to stating their precise definitions.

The expression ``under the same conditions'' deserves a comment.
Clearly we expect that two strictly identical copies of a system
evolving under exactly identical conditions will produce the same
signal $a(\rr,t)$.  The ``same conditions'' must be understood in a
statistical way: the system is prepared by choosing a random initial
configuration extracted from a well-defined probability distribution,
or two identical copies evolve with a random dynamics with known
statistical properties (e.g.\ coupled to a thermal bath at given
temperature and pressure).  We are excluding from consideration cases
where the fluctuations are mainly due to experimental errors.  If that
where the only source of noise, one could in principle repeat the
measurement enough times so that the average $\mean{ a(\rr,t)}$ is
known with enough precision.  Then $\mean{ a(\rr,t)}$ would be an
accurate description of the system's actual spatial and temporal
variations, and the correlations we are about to study would be
dominated by properties of the measurement process rather than by the
dynamics of the system itself.  Instead we are interested in the
opposite case: experimental error is negligible, and the fluctuations
of the observable are due to some process intrinsic to the system.
Indeed in many cases (such as macroscopic observables of systems in
thermodynamic equilibrium) the average of the signal is uninteresting
(it's a constant), but the correlation functions unveil interesting
details of the system's spatial structure and dynamics.

From this discussion it is clear that $a(\rr,t)$ is not an ordinary
function.  Rather, at each particular value of its arguments
$a(\rr,t)$ is a random variable: $a(\rr,t)$ is therefore treated
mathematically as a family of random variables indexed by $\rr$ and
$t$, called \emph{stochastic process} or  \emph{random
  field.}  We don not need to delve into them here (but see
Appendix~\ref{sec:stochastic-processes} for a brief introduction).  To
define correlation functions we only need to assume that the joint
distributions
\begin{equation}
  P_2(a_1,\rr_1,t_1,a_2,\rr_2,t_2) \label{eq:two-point-dist}
\end{equation}
are defined for all possible values of $\rr_1$, $t_1$, $\rr_2$, and
$t_2$.  These distributions characterise only partially the stochastic
process, but they are enough to define the correlation functions we
consider here.

\subsection{Definition of space-time correlation functions}
\label{sec:time-correlations}

The correlation function is the correlation of the random variables
$a(\rr_0,t_0)$ and $a(\rr_0+\rr,t_0+t)$,
\begin{equation}
  \label{eq:def-timecorr}
  C(\rr_0,t_0,\rr,t) = \langle a^*(\rr_0,t_0) a(\rr_0+\rr,t_0+t) \rangle  =   \int\!\!da_1\,da_2\, 
 P(a_1,\rr_0,t_0,a_2,\rr_0+\rr,t_0+t) a_1^* a_2,
\end{equation}
where the star stands for complex conjugate.  The time difference $t$
is sometimes called \emph{lag,} and the function is called \emph{self
  correlation} or \emph{autocorrelation} to emphasize the fact that it
is the correlation of the same observable quantity measured at
different points.  However that from the point of view of probability
theory $a(\rr_0,t_0)$ and $a(\rr_0+\rr,t_0+t)$ are two
\emph{different} (though not independent) random variables.  The
\emph{cross-correlations} of two observables is similarly defined:
\begin{equation}
  \label{eq:6}
  C_{ab}(\rr_0,t_0,\rr,t) = \mean{ a^*(\rr_0,t_0) b(\rr_0+\rr,t_0+t) }.
\end{equation}
The star again indicates complex conjugate.  From now on however we
shall restrict ourselves to real quantities and omit it in the
following equations.

It is often useful to concentrate on the \emph{fluctuation}
$\delta a(\rr,t)=a(\rr,t)-\mean{a(\rr,t)}$, especially when the
average is independent of position and/or time.  The correlation of
the fluctuations is
\begin{equation}
  \label{eq:def-connected}
   C_c(\rr_0,t_0,\rr,t) = \mean{ \delta a(\rr_0,t_0) \delta
     a(\rr_0+\rr,t_0+t) } = 
   C(\rr_0,t_0,\rr,t) - \mean{a(\rr_0,t_0)} \mean{a(\rr_0+\rr,t_0+t)},
\end{equation}
and is called the \emph{connected} correlation function \citep[in
diagrammatic perturbation theory, the connected correlation is
obtained as the sum of connected Feynman diagrams only, see
e.g.][ch.~8]{Binney1992}.  From (\ref{eq:uncorrelated-def}) it
is clear that this function is zero when the variables are
uncorrelated.  For this reason it is often more useful than the
correlation (\ref{eq:def-timecorr}), which for uncorrelated variables
takes the value $\mean{a(\rr_0,t_0)}\mean{a(\rr_0+\rr,t_0+t)}$.

Sometimes a normalised connected correlation is defined so that its
absolute value is bounded by 1, in analogy with the Pearson coefficient:
\begin{equation}
  \rho(\rr_0,t_0,\rr,t) = \frac{C_c(\rr_0,t_0,\rr,t)}
  {\sqrt{C_c(\rr_0,t_0,0,0) C_c(\rr_0+\rr,t_0+t,0,0) }}.
  \label{eq:autocorr-general}
\end{equation}

The names employed here are usual in the physics literature
\citep[e.g.][]{Binney1992,Hansen2005}.  In the mathematical statistics
literature, the connected correlation function is called
\emph{autocovariance} (in fact it is just the covariance of
$a(\rr_0,t_0)$ and $a(\rr_0+\rr,t_0+t)$), while the name
autocorrelation is reserved for the normalized connected correlation
\eqref{eq:autocorr-general}.

The correlations defined here are also known as \emph{two-point}
correlation functions, to distinguish them from higher-order
correlations that can be studied but are outside the scope of this
paper.  Higher-order correlation functions are higher-order moments of
$a(\rr,t)$, while higher-order connected correlations correspond to
the \emph{cumulants} of $a(\rr,t)$ \citep[Ch.~7]{itzykson1989a}
\citep[the distinction between cumulants and moments is only relevant
beyond third order, see e.g.][\S2.2]{van_kampen_stochastic_2007}.

\subsubsection{Global and static quantities}
\label{sec:glob-time-indep}

The full spatial and temporal evolution may be not accessible or not
relevant (e.g.\ because time evolution is very slow and the quantity
is static at the experimental time-scales).  In those cases one
defines static space correlation functions or global time correlation
functions which can be obtained as particular cases
of~\eqref{eq:def-timecorr} and~\eqref{eq:def-connected}.

The static space correlation function is just the space-time
correlation evaluated at $t=0$,
\begin{equation}
  \label{eq:1}
  C(\rr_0,\rr) = C(\rr_0,t_0,\rr,0) =
  \mean{a(\rr_0,t_0) a(\rr_0+\rr,t_0)},
\end{equation}
and analogously for the connected correlation.  Time-only correlation
functions are defined from a signal that can be regarded as a space
integral of a local quantity,
\begin{equation}
  A(t) = \int_V a(\rr,t)\,d\rr,
\end{equation}
so that the time correlation function is
\begin{equation}
  \label{eq:3}
  C(t,t_0) = \mean{A(t_0) A(t_0+t)} = 
  \mean{ \int\!\!d\rr_0 a(\rr_0,t_0) \int\!\! d\rr_1 a(\rr_1,t_0+t) } =
\int \!\! d\rr_0\,d\rr\, C(\rr_0,t_0,\rr,t) .
\end{equation}

\subsubsection{Symmetries}
\label{sec:symmetries}

Knowledge of the symmetries of the system under study, which are
reflected in invariances of the stochastic process used to represent
it, is very important.  Apart from their significance in our
theoretical understanding, at the practical level symmetries are
useful in the estimation of statistical quantities, because they
afford a way of obtaining additional sampling of the stochastic
signal.  For example, if the system is translation-invariant, one can
perform measurements at different places of the same system and treat
them as different samples of the same process, because one knows the
statistical properties are independent of position
(\S\ref{sec:comp-time-corr}).

If a system is time-translation invariant, then
\begin{subequations}
  \begin{align}
    \mean{a(\rr,t)} &= \mean{a(\rr,t+s)},  \\
    \mean{ a(\rr_1,t_1) a(\rr_2,t_2) } &=
                                         \mean{ a(\rr_1,t_1+s) a(\rr_2,t_2+s) }
  \end{align}
 \label{eq:stat-order-2}
\end{subequations}
for all $\rr_1$, $\rr_2$, $t_1$, $t_2$ and $s$.  In this case the
system is said to be \emph{stationary}.

In stationary systems the time origin is irrelevant: no matter
\emph{when} one performs an experiment, the statistical properties of
the observed signal are always the same.  In particular, this implies
that the average $\mean{a(\rr,t)}$ is independent of time (but does
not mean that time correlations are trivial).  Setting $s=-t_1$ one
sees that the second moment depends only on the time difference
$t_2-t_1$, so that the time correlation depends only on one time (in
our definition, the second time, or lag),
\begin{align}
  C(t) &= \mean{A(0) A(t)}, & \text{(stationary)}\\
\intertext{and differs from the connected correlation by a constant:}
  C_c(t) &= \mean{ A(0)A(t)} - \mean{ A}^2 
   =  C(t) -\mean{ A }^2,  & \text{(stationary).}
\label{eq:timecorr-stationary}
\end{align}
This is the situation that holds for a system in thermodynamic
equilibrium.

Translation invariance in space similarly means that
\begin{subequations}
  \begin{align}
    \mean{a(\rr,t)} &= \mean{a(\rr+\ss,t)},  \\
    \mean{ a(\rr_1,t_1) a(\rr_2,t_2) } &=
                                         \mean{ a(\rr_1+\ss,t_1) a(\rr_2+\ss,t_2) } \label{eq:homog}
  \end{align}
\end{subequations}
for all positions and $\ss$.  In a translation-invariant, or
\emph{homogeneous} system, the space origin is irrelevant: no matter
\emph{where} the experiment is performed, the statistical properties
are the same.  In particular $\mean{a(\rr,t}$ is independent of $\rr$,
and the second moments depend only on the displacement $\rr_2-\rr_1$.
Space correlations then depend only on the second argument:
\begin{align}
  C(\rr) &= \mean{a(0,t_0) a(\rr,t_0)},  & \text{(homogeneous)} \\
  C_c(\rr) &= \mean{\delta a(0,t_0) \delta a(\rr,t_0)} = \mean{a(0,t_0) a(\rr,t_0)} - \mean{a(0,t_0)}^2
   =  C(\rr) -\mean{ a(t_0)}^2,  & \text{(homogeneous).}
\end{align}

If in addition the system is \emph{isotropic,} then the process will
be \emph{rotation invariant,} and the second moment depends only on
the distance $r=\lvert \rr\rvert$ between the two points, so
that $C(\rr) = C(r \nn)$ for any unit vector $\nn$.

For stationary and homogeneous systems, the normalised correlation can
be obtained simply by dividing by its value at the origin,
\begin{align}
 \rho(\rr,t) &= \frac{C_c(\rr,t)}{C_c(0,0)}, & \text{(stationary, homogeneous).}  
  \label{eq:autocorr-math-stat-homog}
\end{align}

\subsubsection{Some properties}
\label{sec:prop-time-corr}

It is easy to see from the definitions that in the stationary and
homogeneous case it holds that
\begin{align}
C_c(0,0) &= \mean{a(\rr_0,t_0)^2}-\mean{a(\rr_0,t_0)}^2 = \text{Var}_a, \\
\lvert C_c(\rr,t) \rvert & \le C_c(0,0)  & \forall\, \rr,  t,
\end{align}
and that the correlation is even in $\rr$ and $t$ if $a(\rr,t)$ is
real-valued.

Also, for sufficiently long lags and distances the variables will become
uncorrelated, so that
\begin{align}
  C(\rr\to\infty,t) & \to \mean{a}^2, &   C_c(\rr\to\infty,t) & \to 0,\\
  C(\rr,t\to\infty) & \to \mean{a}^2, & C_c(\rr,t\to\infty) & \to 0
\end{align}
Thus the connected correlation will have a Fourier transform
(\S\ref{sec:fourier-space}).

\subsubsection{Example}
\label{sec:examples}

\begin{figure}
\includegraphics{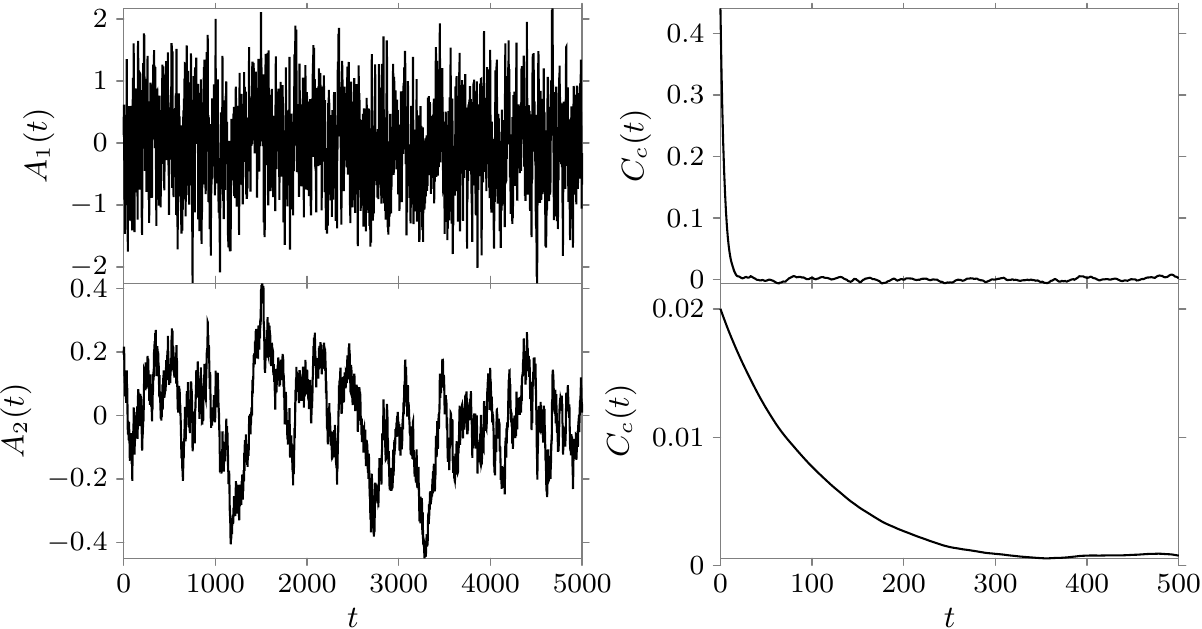}
\caption{Two stochastic signals (left) and their respective connected
  time correlations (right).  Correlation times are $\tau\approx 4.5$
  (top), $\tau\approx100$ (bottom).  The signals were generated
  through Eq.~\eqref{eq:test-sequence} with $\mu=0$, $\sigma^2=2$,
  $N=10^5$, and $w=0.8$ (top), $w=0.99$ (bottom).}
\label{fig:corr-example}
\end{figure}

Let us conclude this section of definitions with an example.
Fig.~\ref{fig:corr-example} shows on the left two synthetic
stochastic signals, generated with the random process described in \S
\ref{sec:example-synth-sign}.  On the right there are the
corresponding connected correlations.  The two signals look different,
and this difference is captured by $C_c(t)$.  We can say that
fluctuations for the lower signal are more persistent: when some value
of the signal is reached, it tends to stay at similar values for
longer, compared to the other signal.  This is reflected in the slower
decay of $C_c(t)$.

\subsection{Correlations in Fourier space}
\label{sec:fourier-space}

Correlation functions are often studied in frequency space, either
because they are obtained experimentally in the frequency domain
(e.g.\ in techniques like dielectric spectroscopy), because the
Fourier transforms are easier to compute or handle analytically, or
because they provide an easier or alternative interpretation of the
fluctuating process.  Although the substance is the same, the precise
definitions used can vary.  One must pay attention to i) the
convention used for the Fourier transform pairs and ii) the exact
variables that are transformed.

Here we define the Fourier transform pairs as
\begin{equation}
  \tilde f(\kk,\omega) = \int_{-\infty}^\infty\!\!dt
  \int\!\! d^dr \, f(\rr,t) e^{-i\kk\cdot\rr + i\omega t}, \qquad
  f(\rr,t) = \int_{-\infty}^\infty \!\frac{d\omega}{2\pi} \int\!\!
  \frac{d^dr}{(2\pi)^d}\, \tilde f(\kk,\omega) e^{i\kk\cdot\rr- i\omega t},
\end{equation}
but some authors choose the opposite sign for the forward transform
and/or a different placement for the $1/2\pi$ factor (sometimes
splitting it between the forward and inverse transforms).  Depending
on the convention a factor $2\pi$ can appear or disappear in some
relations.

Let us consider time-only correlations first.  These are defined in
terms of two times, $t_1$ and $t_2$ (which we chose to write as $t_0$
and $t_0+t$).  One can transform one or both of $t_1$ and $t_2$ or
$t_0$ and $t$.  Let us mention two different choices, useful in
different circumstances.  First, take the connected time correlation
\eqref{eq:def-connected} and do a Fourier transform on $t$:
\begin{equation}
  C_c(t_0,\omega) = \int\!\!dt\,e^{i\omega t} C_c(t_0,t_0+t) =
  e^{-i\omega t_0} \mean{\delta A(t_0) \delta \tilde A(\omega)},
\end{equation}
where $\delta \tilde A(\omega)$ stands for the Fourier transform of
$\delta A(t)$.  This definition is convenient when there is an
explicit dependence on $t_0$ but the evolution with $t_0$ is slow, as
in the physical aging of glassy systems \citep[see
e.g.][]{cugliandolo_dynamics_2004}; one then studies a time-dependent
spectrum.  If on the other hand $C_c$ is stationary, it is more
convenient to write $t_1=t_0$, $t_2=t_0+t$ and do a double transform
in $t_1$, $t_2$:
\begin{multline}
  \tilde C_c(\omega_1,\omega_2) =  \int\!\!dt_1\,dt_2\,e^{i\omega_1
  t_1}e^{i\omega_2 t_2} C_c(t_1,t_2-t_1) = \\
  \int\!\!dt_1\,dt_2\,e^{i\omega_1 t_1}e^{i\omega_2 t_2} \mean{\delta
    A(t_1) \delta A(t_2)} = \mean{\delta\tilde A(\omega_1)\delta\tilde
    A(\omega_2)} =\\
  \int\!\!dt_1\,dt_2\,  e^{i(\omega_1+\omega_2) t_1} e^{i\omega_2 (t_2-t_1)} C_c(t_2-t_1) 
                          = (2\pi) \delta(\omega_1+\omega_2) \tilde C_c(\omega_2),
  \label{eq:fourier-transform-timecorr-double}
\end{multline}
where $\tilde C_c(\omega)$ is the Fourier transform of the stationary
connected correlation with respect to $t$ and we have used the
integral representation of Dirac's delta,
$\delta(\omega-\omega')=(2\pi)^{-1}\int_{-\infty}^\infty e^{i
  t(\omega-\omega')}\,dt$.
As~\eqref{eq:fourier-transform-timecorr-double} shows, the transform
is zero unless $\omega_1=-\omega_2$, so that it is useful to define
the \emph{reduced} correlation,
\begin{equation}
  C_c^R(\omega) = \mean{ \delta\tilde A(-\omega) \delta\tilde
                  A(\omega) } = \mean{ \delta\tilde A^*(\omega)
                  \delta\tilde A(\omega) },
\end{equation}
where the rightmost equality holds when $A$ is real.

The transform $\tilde C_c(\omega)$ is a well-defined function, because
the connected correlation decays to zero (and usually fast enough).
We can then say
\begin{equation}
  \label{eq:redu-almost-eq}
  \tilde C_c(\omega) = \frac{1}{2\pi\delta(0)} C_c^R(\omega).
\end{equation}
This relation can be exploited to build a very fast algorithm to
compute $C_c(t)$ in practice (App.~\ref{sec:algor-comp-time}). The at
first sight baffling infinite factor relating the reduced correlation
to $\tilde C_c(\omega)$ originates in the fact that
$\delta\tilde A(\omega)$ cannot exist as an ordinary function, since
we have assumed that $A(t)$ is stationary.  This implies that the
signal has an infinite duration, and that, its statistical properties
being always the same, it cannot decay to zero for long times as
required for its Fourier transform to exist.  The Dirac delta can be
understood as originating from a limiting process where one considers
signals of duration $T$ (with suitable, usually periodic, boundary
conditions) and then takes $T\to\infty$.  Then
$2\pi \delta(\omega=0) = \int\!\!dt\,e^{i t \omega}|_{\omega=0} =
\int\!\!dt = T$.  These considerations can be made rigorous by
defining the signal's \emph{power spectrum,}
\begin{equation}
  \label{eq:power-spec}
  h(\omega)=\lim_{T\to\infty}\frac{1}{T}\mean{A_T(\omega)A_T^*(\omega)},
  \qquad A_T(\omega) = \int_{-T/2}^{T/2} A(t) e^{i\omega t} \,dt,
\end{equation}
and then proving that $h(\omega)=\tilde C_c(\omega)$ \cite[\S4.7,
\S4.8]{Priestley1981}.

The same considerations apply to space or space-time correlations.
For reference, the relations one finds are
\begin{subequations}
  \begin{align}
    \tilde C_c(\kk_1,\omega_1,\kk_2,\omega_2) &=
                                                \int\!\!d^dr_1 \, dt_1 \,d^dr_2
                                                \,dt_2 \,
                                                e^{-i\kk_1\cdot\rr_1 + i\omega_1 t_1}
                                                e^{-i\kk_2\cdot\rr_2 +
                                                i\omega_2 t_2}
                                                C_c(\rr_1,t_1,\rr_2-\rr_1,t_2-t_1),
    \\ 
    \tilde C_c(\kk,\omega) &= \int_{-\infty}^\infty\!\!dt
                             \int\!\! d^dr \, e^{-i\kk\cdot\rr + i\omega t} C_c(\rr,t), \\
    C_c^R(\kk,\omega) &= \mean{ \delta a(-\kk,-\omega) \delta
                        a(\kk,\omega) }, \\
    \tilde C_c(\kk_1,\omega_1,\kk_2,\omega_2) &= (2\pi)^{d+1}
                                                \delta^{(d)}(\kk_1+\kk_2)
                                                \delta(\omega_1+\omega_2)
                                                C_c^R(\kk_1,\omega_1), \\
    \tilde C_c(\kk,\omega) &= \frac{1}{(2\pi)^{d+1} \delta^{(d)}(0)
                             \delta(0)} C_c^R(\kk,\omega) =
                             \frac{1}{(2\pi)^{d+1} V T} C_c^R(\kk,\omega).
  \end{align}
\end{subequations}

\section{Computing correlation functions from experimental data}

\label{sec:comp-corr-funct}

The theoretical definitions of correlation functions are given as of
averages over some probability distribution, or ensemble.  To compute
them in practice, given data recorded in an experiment or produced in
numerical simulation, there are two issues two consider.  First,
experimental data will be discrete in time as well as in space, either
as a result of sampling, or because the spatial resolution is high
enough to measure the actual discrete units that make up our system
(``particles'', i.e.\ birds, cells, neurons, etc.).  Second, we do not
have direct access to the probability distribution, but only to a set
of samples, i.e.\ results from experiments, distributed randomly
according to an unknown distribution.  Thus we what we actually
compute are \emph{estimators} (as they are called in statistics) of
the averages we want (the correlation functions).  We will discuss
estimators corresponding to different situations, and for clarity it
will be convenient to consider first the case of a global quantity
(correlation in time only, \S \ref{sec:comp-time-corr}) and only later
introduce the additional complications brought in by the space
structure (\S \ref{sec:estim-space-corr}).

Before moving to the estimation of correlations, let us examine the
estimator for the mean, which we will need to estimate the connected
correlation functions, and which will allow us to make a couple of
general points.  It is clearly hopeless to attempt to estimate an
ensemble average unless it is possible to obtain many samples under
the same conditions (i.e., if one is throwing dice, one should throw
many times the \emph{same} dice).  So we assume that we are given a
set of $M$ samples $\{\sampn{a}{n}\}$, $n=1,\ldots,M$ obtained in $M$
equivalent experiments.  From these samples we can estimate the mean
as
\begin{equation}
\overline{a}= \frac{1}{M} \sum_{n=1}^{M} \sampn{a}{n}.
\label{eq:mean-estimate-general}
\end{equation}
This well-known estimator has the desirable property that it ``gets
better and better'' when the number of samples $M$ grows, if the
ensemble has finite variance.  More precisely,
$\overline{a} \to \mean{a}$ as $M\to\infty$.  This property is called
\emph{consistency}, and is guaranteed because
$\mean{\overline{a}}=\mean{a}$ (i.e.\ the estimator is
\emph{unbiased)} and the variance of $\overline{a}$ vanishes for
$M\to\infty$ \citep[\S5.2, \S5.3]{Priestley1981}.

How far off the estimate can be expected to be from the actual value
(i.e.\ the ``error''), is proportional to its variance\footnote{Being
  a function of random variables, the estimator has a probability
  distribution of its own, and in particular mean and variance.}.  A
result called the Cramer-Rao inequality implies that there is a lower
bound to the variance of any estimator, given the number of samples
\citep[\S5.2]{Priestley1981}.  So in practice one of course chooses a
good known estimator such as \eqref{eq:mean-estimate-general}, but the
only way to improve on the error is to acquire more samples, since the
variance of~\eqref{eq:mean-estimate-general} goes as
$\sim 1/\sqrt{M}$.  Unfortunately $1/\sqrt{M}$ is not a very
fast-decreasing function (a tenfold increase in the number of samples
reduces the error only by about a third), so one is always pressed to
obtain the largest possible number of samples.  But in experiments
(especially in biology) the number of samples may be scarce, and more
so in the case of correlations, which are defined in terms of pairs of
values, separated by a fixed distance or time.

To make the most out of the available samples, one must take advantage
of the known symmetries of the system.  Space translation invariance,
allows us to treat samples measured at different points of a large
sample as different equivalent experiments, and the same goes for
samples obtained at different points in time (even if the experiment
has not been restarted) if the system is stationary.  The samples
obtained this way are in general \emph{not} independent (they will be
in fact correlated), but the estimate \eqref{eq:mean-estimate-general}
does not need to be used on independent samples.  It is still unbiased
and consistent, only that its variance will be larger than it would be
for the same number of independent samples (see
Eq.~\eqref{eq:variance-correlated-est}).

Accordingly, whenever in what follows we use $n$ to index samples or
experiments, it is understood that the averages over $n$ can actually be
implemented as averages over a time or space index if the system is
stationary or homogeneous, respectively.  We refer then to \emph{time
  averages} or \emph{space averages} as replacing ensemble averages
when building an estimator.

\subsection{Estimation of time correlation functions}
\label{sec:comp-time-corr}

Assume that the experiment records a scalar signal with a uniform
sampling interval $\Delta t$, so that we are handled $N$ real-valued
and time-ordered values forming a sequence $A_i$, with
\(i=1,\ldots, N\).  It is understood that if the data are digitally
sampled from a continuous time process, proper filtering has been
applied\footnote{According to the Nyquist-Shannon sampling theorem, if
  the signal has frequency components higher than half the sampling
  frequency (i.e.\ if the Fourier transform is nonzero for
  $\omega\ge \pi /\Delta t$) then the signal cannot be accurately
  reconstructed from the discrete samples; in particular the high
  frequencies will ``polute'' the low frequencies (an effect called
  aliasing).  Thus the signal should be passed through an analog
  low-pass filter before sampling.  See \citet[\S 12.1]{Press1992a}
  for a concise self-contained discussion, or \citet[\S
  7.1]{Priestley1981}.}.  In what follows we shall measure time in
units of the sampling interval, so that in the formulae below we shall
make no use of $\Delta t$.  To recover the original time units one
must simply remember that $A_i = A(t_i)$ with
$t_i=t_0 + (i-1) \Delta t$.  For the stationary case we shall write
$C_k=C(t_k)$ where $t_k$ is the time difference, $t_k= k \Delta t$ and
$k=0,\ldots,N-1$, and in the non-stationary case $C_{i,k}=C(t_i,t_k)$.

If the process is not stationary, nothing can be estimated from a
single sequence because the successive values correspond to different,
nonequivalent experiments, as the conditions have changed from one
sample to another (i.e.\ the system has evolved in a way that has
altered the characteristic of the stochastic process).  In this case
the correlation depends on two times and the mean itself can depend on
time.  The only way to estimate the mean or the correlation function
is to obtain many sequences $A_i^{(n)}$, $n=1,\ldots,M$ reinitializing
the system to the same macroscopic conditions each time (in a
simulation, one can for example restart the simulation with the same
parameters but changing the random number seed).  It is not possible
to resort to time averaging, and the estimates are obtained by
replacing the ensemble average by averages across the different
sequences, i.e.\ the (time-dependent) mean is estimated as
\begin{equation}
  \mean{A(t_i)} \approx \overline{A_i} =\frac{1}{M} \sum_{n=1}^M A_i^{(n)},
  \label{eq:est-mean-nonstationary}
\end{equation}
and the time correlation as
\begin{equation}
  C_c(t_i,t_k) \approx \CC_{i,k} =\frac{1}{M}\sum_n^M \delta A_i^{(n)}
  \delta A_{i+k}^{(n)}  , \qquad \delta A_i^{(n)} = A_i^{(n)} - \overline{A_i},
  \label{eq:conn-corr-ns-est}
\end{equation}
where the hat distinguishes the estimate from the actual quantity.
Both estimators are consistent and unbiased, i.e.\
$\CC_{i,k} \to C(t_i,t_i)$ and $\overline{A(t_i)}\to \mean{A(t_i)}$
for $M\to\infty$.

If the system is stationary, one can resort to time averaging to build
estimators (like \eqref{eq:conn-corr-estimate} below) that only
require one sequence.  But let us remark that it is always
sensible to check whether the sequence is actually stationary.  A
first check on the mean can be done dividing the sequence in several
sections and computing the average of each section, then looking for a
possible systematic trend.  If this check passes, then one should
compute the time correlation of each section and then checking that
all of them show the same decay (using fewer sections than in the
first check, as longer sections will be needed to obtain meaningful
estimates of the correlation function).  It is important to note that
this second check is necessary even if the first one passes; as we it
is possible for the mean to be time-independent (or its
time-dependence undetectable) while the time correlation still depends
on two times.

In the stationary case, then, we can estimate the mean with
\begin{equation}
  \overline{A} = \frac{1}{N}\sum_{i=1}^N A_i,   \label{eq:mean-stationary-estimate}
\end{equation}
and the connected correlation function with
\begin{equation}
  \CC_k = \frac{1}{N-k} \sum_{j=1}^{N-k} \delta A_j \delta A_{j+k},
  \qquad
  \delta A_j = A_j - \overline{A}. \label{eq:conn-corr-estimate}
\end{equation}
In the last equation, the true sample mean $\mean{A}$, if known, can
be used instead of the estimate $\overline{A}$.  If the true mean is
used, the estimator \eqref{eq:conn-corr-estimate} is unbiased,
otherwise it is \emph{asymptotically} unbiased, i.e.\ the bias tends
to zero for $N\to\infty$, provided the Fourier transform of $C_c(t)$
exists.  More precisely,
$\langle \CC_k \rangle = C_c(t_k) - \alpha/N$, where
$\alpha=2\pi \text{Var}_A \tilde C_c(\omega=0)$.  The \emph{variance}
of $\CC_k$ is $O(\frac{1}{N-k})$ \citep[\S5.3]{Priestley1981}.  This
is sufficient to show that, at fixed $k$, the estimator is consistent,
i.e. $\CC_k\to C_c(t_k)$ for $N\to\infty$.  However, the variance
grows with increasing $k$, and thus the tail of $\CC_k$ is
considerably noisy.  In practice, for $k$ near $N$ the estimate is
mostly useless, and the usual rule of thumb is to use $\CC_k$ only for
$k\le N/2$.

Knowing the time correlation, one can make the statement that the
variance of the estimate \eqref{eq:mean-stationary-estimate} is larger
than that of \eqref{eq:est-mean-nonstationary} quantitative: the
variance of the estimate with correlated samples is
\cite{sokal_monte_1997}
\begin{equation}
  \text{Var}_{\overline{a}} \approx \frac{2\tauint}{N} \left [
    \mean{A^2}-\mean{A}^2 \right],
  \label{eq:variance-correlated-est}
\end{equation}
i.e.\ $2\tauint$ times larger than the variance in the independent
sample case, where $\tauint$ is the integral correlation time defined
below \eqref{eq:tauint}.  In this sense $N/2\tauint$ can be thought of
as the number of ``effectively independent'' samples.

We have not written an estimator for the nonconnected correlation
function.  It is natural to propose
\begin{equation}
  \hat C_k  = \frac{1}{N-k} \sum_{j=1}^{N-k} A_j
             A_{j+k}.   \label{eq:corr-estimate} 
\end{equation}
However, although $\hat C_k$ is unbiased, it can have a large variance
when the signal has a mean value larger than the typical fluctuation
(see \S\ref{sec:two-prop-estim}).  As a general rule, it is not a good
idea estimate the connected correlation by using
\eqref{eq:corr-estimate} and then subtracting $\overline{A}^2$.
Instead, the connected estimator \eqref{eq:conn-corr-estimate} should
be used.  An possible exception may be the case when an experiment can
only furnish many short independent samples of the signal (see the
examples in \S\ref{sec:example-synth-sign}).

If several sequences sampled from a stationary system are available, it
is possible to combine the two averages: one \emph{first} computes the
stationary correlation estimate \eqref{eq:conn-corr-estimate} for each
sequence, and \emph{then} averages the different correlation estimates
(over $n$ at fixed lag $k$).  It is clearly wrong to average the
sequences themselves before computing the correlation, as this will
tend to approach the (stationary) ensemble mean $\mean{A}$ for all
times, destroying the dynamical correlations.

There is another asymptotically unbiased estimator of the connected
correlation, that can be obtained by using $1/N$ instead of $1/(N-i)$
as prefactor of the sum in \eqref{eq:conn-corr-estimate}.  Calling
this estimator $C_{c,k}^*$, it holds that
$\langle C_{c,k}^*\rangle = C_c(t_k) - \alpha/N - k C(t_k)/N - \alpha
k/N^2$, where $\alpha$ is defined as before, and again $\alpha=0$ if
the exact sample mean is used \citep[\S 5.3]{Priestley1981}.  This has
the unpleasant property that the bias depends on $k$, while the bias
of $\hat C_{c,k}$ is independent of its argument, and smaller in
magnitude.  The advantage of $C_{c,k}^*$ is that its variance is
$O(1/N)$ independent of $k$, thus it has a less noisy tail.  Some
authors prefer $C_{c,k}^*$ due to its smaller variance and to the fact
that it strictly preserves properties of the correlation, which may
not hold exactly for $\hat C_{c,k}$.  Here we stick to $\hat C_{c,k}$,
as usual in the physics literature
\citep[e.g.][]{sokal_monte_1997,Allen1987,Newman1999}, so that we
avoid worrying about possible distortions of the shape.  In practice
however, it seems that as long as $N$ is greater than $\sim 10\tau$ (a
necessary requirement in any case, see below), and for $k\le N/2$,
there is little numerical difference between the two estimates.

\subsubsection{Two properties of the estimator}
\label{sec:two-prop-estim}

We must mention two important features of the estimator that are
highly relevant when attempting to compute numerically the time
correlation.  The first is that the difference between the
non-connected and connected estimators is not ${\overline a}^2$, but
\begin{equation}
 \hat C_k - \hat C_{c,k} = \overline{a} \left[ \frac{1}{N-k}
\sum_j^{N-k} a_j + \frac{1}{N-k}\sum_j^{N-k} a_{j+k}
-\overline{a} \right],
\end{equation}
as it is not difficult to compute.  The difference does tend to
\(\overline{a}^2\) for \(N\to\infty\), but in a finite sample
fluctuations are important, especially at large $k$.  Since
fluctuations are amplified by a factor $\overline a$, when the signal
mean is large with respect to its variance, the estimate $\hat C_k$ is
completely washed out by statistical fluctuations.  This is why, while
$C(t)$ and $C_c(t)$ differ by a constant, in practice it is a bad idea
to compute $\hat C_k$ and subtract $\overline{a}^2$ to obtain an
estimate of the connected correlation.  Instead, \emph{one computes
  the connected correlation directly} by estimating first the mean and
then using~\eqref{eq:conn-corr-estimate}.

Another important fact is that the estimator of the connected
correlation \emph{will always have zero,} whatever the length of the
sample used, \emph{even if $N$ is much shorter than the correlation
  time.}  As shown in App.~\ref{sec:zero-cross}, for any time series
one has that $\CC_0>0$ and that $\CC_{c,k}$ will \emph{change sign} at
least once for $k\ge1$ (this applies to the case when the mean and
connected correlation are estimated with a single time series).  The
practical consequence is that when $N$ is of the order of $\tau$, or
smaller, the estimate $\CC_k$ will suffer from strong finite-size
effects, and its shape will be quite different from the actual
$C_c(t)$.  In particular, since $\CC_k$ will intersect the $x$-axis,
it will look like the series has decorrelated when in fact it is still
highly correlated.  Be suspicious if $\CC_k$ changes sign once and
stays very anticorrelated.  Anticorrelation may be a feature of the
actual $C_c(t)$, but if the sample is long enough, the estimate should
decay until correlation is lost (noisily oscillating around zero).
One must always perform tests estimating the correlation with
different values of $N$: if the shape of the correlation at short
times depends on $N$, then $N$ must be increased until one finds that
estimates for different values of $N$ coincide for lags up to a few
times the correlation time (we show an example next).  Once a
sample-size-independent estimate has been obtained, the correlation
time can be estimated (\S\ref{sec:correlation-time}), and it must be
checked that self-consistently $N$ is several times larger than
$\tau$.

\subsubsection{Example}
\label{sec:example-synth-sign}

To illustrate the above considerations on estimating $C_c(t)$ we use a
synthetic correlated sequence generated from the recursion
\begin{equation}
 a_i = w a_{i-1} + (1-w) \xi_i,  \label{eq:test-sequence}
\end{equation}
where $w\in[0,1]$ is a parameter and the $\xi_i$ are uncorrelated
Gaussian random variables with mean $\mu$ and variance $\sigma^2$.
Assuming the stationary state has been reached it is not difficult to
find
\begin{equation}
  \mean{a}=\mu, \qquad \mean{(a-\mu)^2}=\frac{1-w}{1+w}\sigma^2, \qquad C^{(c)}_k
  =\mean{(a_i-\mu)(a_{i+k}-\mu)} = \sigma^2 w^k,
\end{equation}
so that the correlation time is $\tau=-1/\log w$.

\begin{figure}
  \centering
  \includegraphics{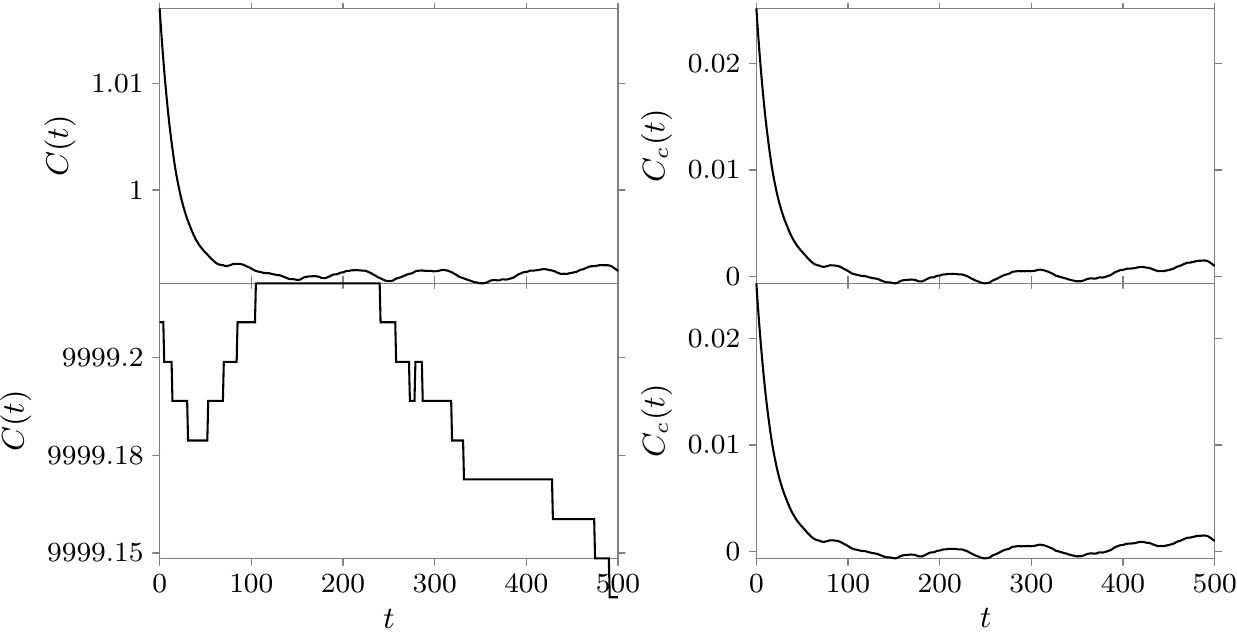}
  \caption{Connected vs.\ nonconnected estimate.  The estimate of
    $C(t)$ (equation \eqref{eq:corr-estimate}, left panels) is much
    worse that the estimates of $C_c(t)$ (equation
    \eqref{eq:conn-corr-estimate}, right panels).  The (artificial)
    signal was generated with \eqref{eq:test-sequence}. Top panels:
    $\mu=1$; bottom panels: $\mu=100$.  In both cases, $\sigma^2=1$,
    $w=0.95$ ($\tau\approx20$) and length $N=5\cdot 10^4$. }
  \label{fig:conn-vs-nonconn}
\end{figure}

We used the above recursion to generate artificial sequences and
computed their time correlation functions with the estimates discussed
above.  Fig.~\ref{fig:conn-vs-nonconn} shows the problem that can face
the non-connected estimate.  When the average of the signal is smaller
than or of the order of the noise amplitude (as in the top panels),
one can get away with using \eqref{eq:corr-estimate}.  However if
$\mu\gg\sigma$, the non-connected estimate is swamped by noise, while
the connected estimate is essentially unaffected (bottom panels).
Hence, if one is considering only one sequence, one should always use
the connected estimator.

\begin{figure}
  \centering
  \includegraphics{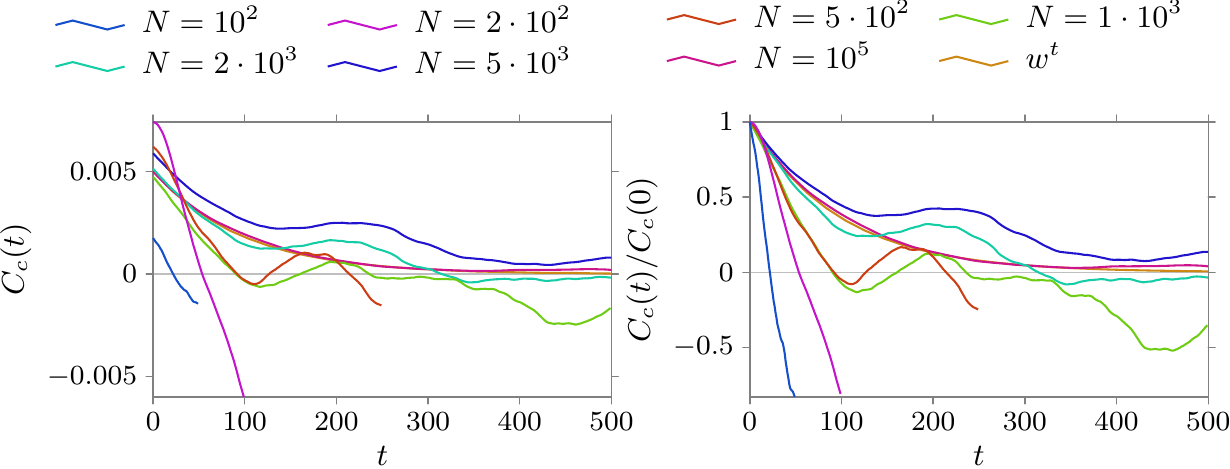}
  \caption{Finite size effects. Estimates of the connected correlation
    (left) and normalized connected correlation (right) for sequence
    \eqref{eq:test-sequence} of different lengths as indicated,
    together with the analytical result $C_c(t)=\sigma^2
    w^t$. Parameters are $\mu=0$, $\sigma^2=1$, $w=0.99$
    ($\tau\approx100$). }
  \label{fig:finite-size}
\end{figure}

In Fig.~\ref{fig:finite-size} we see how using samples that are too
short affects the correlation estimates.  The same artificial signal
was generated with different lengths.  For the shorter lengths, it is
seen that the correlation estimate crosses the $t$ axis (as we have
shown it must) but does not show signs of losing correlation.  One
might hope that $N=1000\approx 10\tau$ is enough (the estimate starts
to show a flat tail), but comparing to the result of doubling the
length shows that it is still suffering from finite-length effects.
For this particular sequence, it is seen that at least $20\tau$ to
$50\tau$ is needed to get the initial part of the normalized connected
correlation more or less right, while a length of about 1000$\,\tau$
is necessary to obtain a good estimate up to $t\sim5\tau$.  The
unnormalized estimator suffers still more from finite size due to the
increased error in the determination of the variance (left panel).

\begin{figure}
  \centering
  \includegraphics{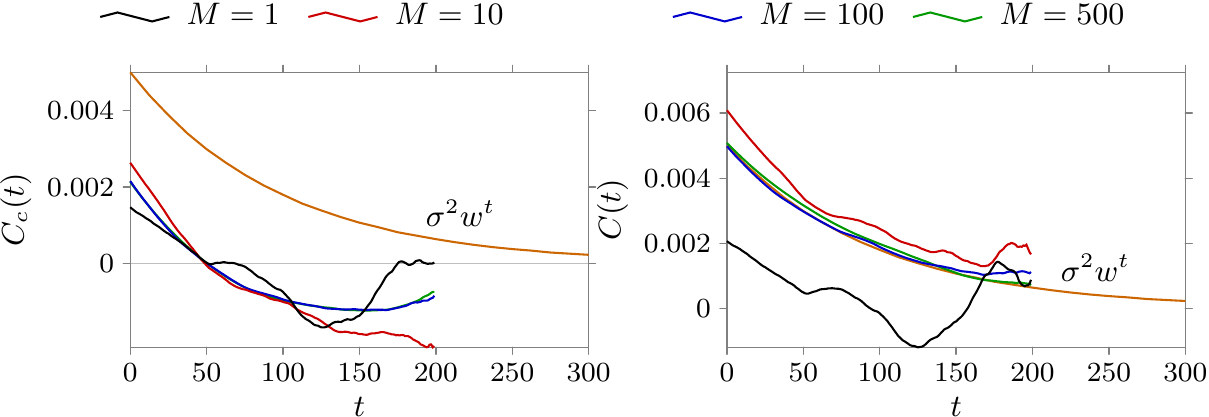}
  \caption{Effect of averaging the estimates of many short sequences.
    We show the connected (left) and non-connected (right) estimates
    of $M$ different samples of sequence \eqref{eq:test-sequence} as
    indicated in the legend, with $N=200$, $\mu=0$, $\sigma^2=1$,
    $w=0.99$ ($\tau\approx100$). The analytical result $C_c(t)=\sigma^2
    w^t$ is also plotted.}
  \label{fig:many-short}
\end{figure}

If the experimental situation is such that it is impossible to obtain
sequences much longer than the correlation time, one can get some
information on the time correlation if it is possible to repeat the
experiment so as to obtain several \emph{independent} and
\emph{statistically equivalent} sample sequences.  In
Fig.~\ref{fig:many-short} we take several sequences with the same
parameters as in the previous example, but quite short
($N=200\approx 2\tau$).  As we have shown, it is not possible to
obtain a moderately reasonable estimate of $C_c(t)$ using one such
sequence (as is also clear from the $M=1$ case of
Fig.~\ref{fig:many-short}).  However, the figure shows how one may
benefit from the multiple repetitions of the (short) experiment by
averaging together all the estimates.  The averaged connected
estimates are always far from the actual correlation function, even
for $M=500$ (the case which contains in total $10^5$ points, which
proved quite satisfactory in the previous example): this is
consequence of the fact that all connected estimates must become
negative.  Instead, the averaged non-connected estimates approach
quite closely the analytical result even though not reaching the
decorrelated region\footnote{Note that in this case $\mu=0$ so that
  fluctuations are larger than the average.  If the average were very
  large, one may attempt to compute a connected correlation estimate
  by using all sequences to estimate the average, then substracting
  this same average to all sequences.}.  Although it is tricky to try
to extract a correlation time from this estimate (due to lack of
information on the last part of the decay), this procedure at least
offers a way to obtain some dynamical information in the face of
experimental limitations.

\subsection{Estimation of space correlation functions}
\label{sec:estim-space-corr}

To compute space correlations one needs experimental data with spatial
resolution.  We can assume that each experiment provides us a set of
values $\{\sampn{a_i}{n}\}$ together with a set of positions
$\{\sampn{\rr_i}{n}\}$ which specify where in the sample the $i$-th
value has been recorded.  The experiment label, $n=1,\ldots,M$ can
indicate experiments on different samples, or for a stationary system,
it can be a time index.

The positions $\sampn{\rr_i}{n}$ may be fixed by the experiment (e.g.\
when sampling in space with an array of sensors, in which case they
are likely independent of $n$) or might be themselves part of the
experimental result (as when measuring the positions of moving
organisms).  In the former case, a naive implementation of
definition~\eqref{eq:1} will pick up the structure of the experimental
array itself, introducing correlations that are clearly an artefact.
In the latter case, the positions may encode nontrivial structural
features, which will show up in the correlation function of $a$
together with $a$-specific effects.  For example, particles with
excluded volume will have non-trivial density correlations at short
range, while there might be long-range correlations due to an ordering
of $a$.  It is then wise to untangle the two effects as much as
possible.  In both cases it is thus convenient to implement the 
definition in a way that separates structural correlations.

Structural effects can be studied through density correlation
functions.  One such function is the \emph{pair distribution function}
  \citep[\S 2.5]{Hansen2005}, defined for homogeneous systems as
\begin{equation}
  g(\rr) = \frac{1}{\rho_0 N} \mean{ \sum_{i\neq j|} \delta^{(d)}(\rr -
    \rr_{ij}) }
  = \frac{1}{N\rho_0} \mean{ \sum_{i\neq j} \frac{ \delta\bigl(\lvert
    \rr\rvert-\lvert r_{ij}\rvert \bigr) }{4\pi \lvert \rr\rvert^2} },
\label{eq:gofr}
\end{equation}
where $\delta^{(d)}(\rr)$ is the $d$-dimensional Dirac's delta,
$\rr_{ij}=\rr_i-\rr_j$ is the pair displacement and $\rho_0 = N / V$
is the average density.  The second equality is valid for isotropic
systems, and $g(r)$ is often called \emph{radial} distribution
function in this case.  Defining the density of the configuration
$\{\rr_i\}$ by $\rho(\rr) = \sum_i \delta(\rr-\rr_i)$, the pair
distribution function is found to be related to the density-density
correlation function by
\begin{equation}
  \mean{\rho(\rr_0)\rho(\rr_0+\rr)} = \rho_0^2 g(\rr) + \rho_0
  \delta(\rr) = \rho_0^2 g(\lvert \rr\rvert) +
  \rho_0\frac{\delta(\lvert \rr\rvert)}{4\pi \lvert \rr\rvert^2},
\end{equation}
where again the second equality applies to isotropic systems.  The
Dirac delta term arises because in $\mean{\rho(\rr_0)\rho(\rr_0+\rr)}$
one includes the $i=j$ terms that were excluded from the sum
in~\eqref{eq:gofr}.  If the $\{\rr_i\}$ are obtained with periodic
boundary conditions (as is often the case in simulation),
\eqref{eq:gofr} can be used to estimate $g(r)$ almost directly: it
suffices to replace Dirac's delta by a discrete version (i.e.\ turn it
into a binning function) and the average by a (normalised) sum over
all available statistically equivalent configurations.  For the
isotropic case,
\begin{subequations}
  \label{eq:gr-estimator}
  \begin{align}
    \hat g_k &= \frac{1}{M} \sum_n \frac{1}{\rho_0 N} \sum_{ij} \frac{\Delta\left[\sampn{r_{ij}}{n} -
               (k+1/2)\Delta r \right]}{V_k},    \\
    \Delta(r) &=
                \begin{cases}
                  1  & \text{if } -\Delta r/2 < r \le \Delta r/2,\\
                  0 & \text{otherwise}.
                \end{cases}
  \end{align}
\end{subequations}
where $\Delta r$ is the bin width and $V_k$ is the volume of the
$k$-th bin, in 3 dimensions
$V_k = \frac{4}{3}\pi (\Delta r)^3\left( (k+1)^3-k^3\right)$.  The bin
width must not be chosen too small or some bins will end up having
very few or no particles.  In any case, it must be larger than the
experimental resolution, otherwise $g(r)$ will pick up spurious
correlations arising from the fact that due to finite spatial
resolution, all positions effectively lay on the sites of a square
lattice.

If borders are non-periodic, as in experimental data, using
\eqref{eq:space-cc-estimate} will introduce bias at large $r$, because
the number of neighbours of a given particle stops growing as $r^2$
(the volume of the bin) as soon as $r$ is of the order of the distance
of the particle to the nearest border.  In this case an unbiased
estimator is
\begin{align}
 \hat g_k &= \frac{1}{\rho_0 N_k} \sum_{i\in S_k} \sum_j \frac{\Delta( k -
             r_{ij}/\Delta r)}{V_k}, 
\label{eq:gr-hanisch}
\end{align}
where $i\in S_k$ means that the sum runs only over the set of
particles that are at least at a distance $(k+1/2)\Delta r$ from the
nearest border, and $N_k$ is the number of such particles.  In other
words, for a given point $i$, the pair $ij$ only enters the sum if
$\rr_j$ lies on a bin (centred at $i$) that is completely contained
within the system's volume. This is known as the unweighted Hanisch
method \citep{hanisch_remarks_1984}.

To conclude our remarks on structural correlation functions, let us
note that if correlations in Fourier space are of interest, it is
typically better to estimate them directly, rather than going through
the real space functions and applying a numerical Fourier transform.
For example, the \emph{structure factor,} which is proportional to the
reduced density correlation,
\begin{equation}
  S(\kk) = \frac{1}{N} \mean{\rho(\kk)\rho(-\kk)} = 1 + \rho_0 \tilde g(\kk),
\end{equation}
where $\tilde g(\kk)$ is the Fourier transform of $g(\rr)$, can be be
computed directly from the $\sampn{\rr_i}{n}$ through
\begin{equation}
  S(\kk) = \frac{1}{M}  \frac{1}{N} \sum_n   \sum_{jl} e^{i \kk\cdot
    \sampn{\rr_{jl}}{n}}  =
   1 + \frac{1}{N M}\sum_n \sum_{i\neq j} \frac{\sin \lvert \kk\rvert \sampn{r_{ij}}{n}}{\lvert
      \kk\rvert \sampn{r_{ij}}{n}} ,
\end{equation}
where again the second equality applies to isotropic system and is
obtained by analytically averaging the first expression over all
angles.  For further discussion of density correlation functions we
refer the reader to condensed matter texts such as \citet{Hansen2005}
or \citet{Ashcroft1976}.

Turning now to the correlations of $a$, and focusing on the
homogeneous case, we seek an estimator of the connected correlation
function.  How to write it depends on exactly what one means by
$a(\rr)$ (it can be defined as a density or as specific quantity,
i.e.\ per unit volume or per particle), and on the fact that one wants
to separate structural correlations.  These issues are discussed at
length in \citet{cavagna_physics_2018}.  Here it will suffice to
state that a good estimator is
\begin{equation}
  \label{eq:space-cc-estimate}
  \CC(\rr) = \frac{\sum_{ij} \delta a_i \delta a_j
    \delta(\rr-\rr_{ij}) } {\sum_{kl} \delta(\rr-\rr_{kl})}
   = \frac{\sum_{ij} \delta a_i \delta a_j
    \delta(r-r_{ij}) } {\sum_{kl} \delta(r-r_{kl})},
\end{equation}
where again $\delta a_i = a_i-\overline{a}$ and the last expression is
for the isotropic case.  We have written $\delta$ functions for
clarity, but clearly one uses binning as in~\eqref{eq:gr-estimator}.
These expressions do a good job of disentangling correlations of the
property $a$ from structural correlations.  In the case of a fixed
lattice, $\CC(\rr)$ is completely emancipated from the effects of the
lattice structure, while in the case of moving particles, at least the
uncorrelated effects of structure will be taken care of.  Also, thanks
to the denominator, \eqref{eq:space-cc-estimate} is free from boundary
effects, so that one does need to worry about the border bias that
requires the use of the Hanisch method for $g(r)$.

As written, \eqref{eq:space-cc-estimate} can be computed using a single
configuration.  Since the system is homogeneous, the mean $\mean{a}$
can also be estimated with a space average,
\begin{equation}
  \label{eq:ave-spaceave}
  \overline{a} = \frac{1}{N} \sum_i a_i .
\end{equation}
We shall write $\CC_\text{sp}(r)$ (for space average) when it is
computed for single configurations.  If several configurations
equivalent and with the same boundary conditions, are available,
additional averaging using these can be performed.  We shall still
call the estimate a space average if the correlation is first computed
for each configurations, with the mean estimated with a  average of
the same configuration, and the average over configurations is done
afterwards.

Another way to proceed is to use all configurations to compute
\emph{first} an estimate of the mean, and then compute the
correlation.  In this case we shall write $\CC_\text{ph}(r)$, for
phase average, because this procedure is closest to actually
performing a phase-space, or ensemble, average.

An important fact about the estimate ~\eqref{eq:space-cc-estimate} in
the space-average case is that $\CC_\text{sp}(\rr)$ will always have a
zero, exactly like the estimator for the connected time
correlation~\eqref{eq:conn-corr-estimate}.  The reason is similar, and
it can be seen considering that since $\sum_i \delta a_i=0$, then
\begin{equation}
  \label{eq:2}
  0 = \frac{1}{N} \sum_{ij}\delta a_i\delta a_j = \rho_0\int\!\!
  d\rr\, g_F(\rr) C(\rr),
\end{equation}
where $g_F(\rr)$ is defined as the pair distribution
function~\eqref{eq:gofr} but without excluding the case $i=j$ from the
double sum.  Since $g_F(r)>0$, it follows that $\CC_\text{sp}(\rr)$
must change sign, so there is always an $r_0$ such that
$\CC_\text{sp}(r_0)=0$.  This is an artefact, but it can be exploited
to obtain a proxy of the correlation scale in critical systems
(\S\ref{sec:finite-size-effects}).

Let us conclude by writing the Fourier-space counterpart of $\CC(r)$.
We introduce it as proportional to the reduced correlation,
\begin{equation}
  \hat C(\kk) = \frac{1}{N} \delta a(\kk)\delta(-\kk) = \frac{1}{N}
  \sum_{ij} \delta a_i \delta a_j e^{i\kk\cdot\rr_{ij}}=
   \frac{1}{N} \sum_{ij} \frac{\sin \lvert \kk \rvert r_{ij}} {\lvert
     \kk\rvert r_{ij}}  \delta  a_i \delta a_j.
  \label{eq:kk-space-estimate}
\end{equation}
It is related to the real space correlation by
\begin{equation}
  \hat C(k) = \rho_0 \int\!\!d\rr\, g_F(r) e^{-i\kk\cdot\rr} \CC(r).
\end{equation}
The function $g_F(r)$ appears in the integral (in effect reintroducing
structural effects in $\hat C(k)$) because
\eqref{eq:kk-space-estimate} is more convenient computational-wise
than the Fourier transform of \eqref{eq:space-cc-estimate}, and is
consistent with an integration measure chosen so that the integral of
$a(r)$ equals $\sum_i a_i$, which is often the global order parameter
\citep[see][App.\ A]{cavagna_physics_2018}.

\subsection{Estimation of space-time correlations}
\label{sec:space-time}

Space-time correlation estimation brings together the issues
encountered in the time and space cases.  The previous discussion
should allow the reader to generalise the estimators written above to
the space-time case.  For brevity, let us just write the space-time
generalisation of~\eqref{eq:space-cc-estimate} and
~\eqref{eq:kk-space-estimate} (for homogeneous and stationary
systems):
\begin{align}
  \CC(\rr,t) & = \frac{1}{T} \sum_{t_0} \frac{\sum_{ij} \delta a_i(t_0) \delta a_j(t_0+t)
    \delta( \rr - \rr_i(t_0) + \rr_j(t_0+t) )}
               {\sum_{ij}   \delta( \rr - \rr_i(t_0) + \rr_j(t_0+t) ) }, \\
  \hat C(\kk) &= \frac{1}{T N} \sum_{t_0} \sum_{ij} \delta a_i(t_0)
                \delta a_j(t_0+t) e^{-i\kk \cdot [\rr_i(t_0) - \rr_j(t_0+t)}.
\end{align}


\section{Correlation length and correlation time}
\label{sec:relax-time-corr}

A connected space correlation function measures how correlation is
gradually lost as one considers two locations further and further
apart. Similarly, the connected time correlation function measures how
correlation is gradually lost as time elapses.  One often seeks for a
summary of this detailed information, in the form of a (space or time)
\emph{scale} that measures the interval it takes for significant
decorrelation to happen: these are the \emph{correlation length} $\xi$
and the \emph{correlation time}\footnote{Sometimes the term relaxation
  time is used as synonymous with correlation time.  Actually, the
  relaxation time is the time scale for the system to return to
  stationarity after an external static perturbation is applied or
  removed.  These two scales are equal for systems in thermodynamic
  equilibrium, as a consequence of the fluctuation-dissipation theorem
  \cite{kubo_statistical_1998}, so the exchange of terms is admissible
  in that case.  However, this is not to be taken for granted in the
  general out-of-equilibrium case.}  $\tau$.

Sometimes these are interpreted as the distance or time after which
the connected correlation has descended past a prescribed level.
Although this can be a useful operational definition when working with
a set of correlation functions of similar shape, these scales are
better understood as the asymptotic exponential rate of the decay.
The precise numerical value (which will depend on the details of the
computation method) is less important than their trend with the
relevant control parameters (temperature, concentration, $\ldots$): in
this sense one can say that $\xi$ and $\tau$ are most useful to
compare the correlation range as some environmental condition is
changed.

Of course, correlation functions are typically more complicated than a
simple exponential, and it may make sense to describe them using more
than one scale (e.g.\ as a superposition of exponential decays).
\emph{The}\/ correlation length refers to the rate of the last
exponential to die out.  More precisely \citep[\S II.2-3]{amit2006}
\begin{equation}
  \xi = \lim_{r\to\infty} \frac{r}{-\log C_c(r)}.
  \label{eq:xi-formaldef}
\end{equation}
This definition, with emphasis on the long-range behaviour, is
particularly suited to the study of systems in the critical region.
Note that $\xi=\infty$ does not mean that the correlation never
decays, but only that it does so slower than exponentially (i.e.\ as a
power law).

Similarly, the correlation time is \citep{sokal_monte_1997},
\begin{equation}
  \tau = \lim_{t\to\infty} \frac{t}{-\log C_c(t)}.
  \label{eq:tau-formaldef}
\end{equation}

These definitions work well when the decay is a combination of
exponentials and power laws.  Exponentials of powers, on the other
hand, never yield a finite scale according to~\eqref{eq:xi-formaldef}
or~\eqref{eq:tau-formaldef}, we defer discussion of those to
App.~\ref{sec:fits-time-corr}.

Clearly definitions~\eqref{eq:xi-formaldef}
and~\eqref{eq:tau-formaldef} are not directly applicable to extract
$\xi$ or $\tau$ from correlation functions computed in experiments or
simulations, which are of finite range.  The rest of this section is
devoted to discussing ways to obtain $\xi$ or $\tau$ from finite data.

A possibility is to fit the decay to some function, and apply the
definitions to this function.  This may be fine, as long as one can
avoid proliferation of parameters\footnote{``With four parameters I
  can fit an elephant, and with five I can make him wiggle his
  trunk.''  Attributed to John von Neumann
  \cite{dyson2004}.}. However, it can be difficult to fit both the
short- and long-range regimes, so an alternative is to attempt fit
$ \log \left[ x^\alpha C_c(x) \right] $ (where $x$ stands for $r$ or
$t$) for large $x$ with straight line $-A x$.  If the fit is good then
one has $\xi=1/A$ or $\tau=1/A$.  One expects $\alpha=0$ for the time
case (at least in equilibrium), and in the space case
$\alpha\approx 0$ in $d=2$ and $\alpha\approx 1$ in $d=3$ according to
the scaling form \eqref{eq:cr-scaling}.  This power has no effect on
the definition \eqref{eq:xi-formaldef}, but is important to obtain a
reasonable fit on relatively short range.

Nevertheless, fitting may not be the best strategy, especially if the
linear region in the fit is short and the slope is very sensitive to
the choice of fitting interval.  We present next some alternative
procedures.

\subsection{Correlation time}
\label{sec:correlation-time}

A quite general way to define a correlation time is from the integral
of the normalized connected correlation,
\begin{equation}
\tauint =  \int_0^\infty \rho(t) \, dt . \label{eq:tauint}
\end{equation}
Clearly for a pure exponential decay $\tauint=\tau$.  In general, if
$\rho(t)=f(t/\tau)$ then $\tauint = \text{const}\,\tau$, so that
$\tauint$ has the same dependence on control parameters as $\tau$.
This would change if $\rho(t) = (t/t_0)^\alpha f(t/\tau)$
\cite[\S2]{sokal_monte_1997}, but one expects $\alpha=0$ in
equilibrium.  In any case, $\tauint$ has a meaning of its own, as the
scale that corrects the variance of the estimate of the mean when
computed with correlated samples (\S\ref{sec:comp-time-corr}).

With some care, $\tauint$ can be computed from experimental or
simulation data, avoiding the difficulties associated with thresholds
or fitting functions.  The following procedure \cite{sokal_monte_1997}
is expected to work well if long enough sequences are at disposal and
the decay does not display strong oscillations or anticorrelation.  If
$\CC_k\approx C_c(k\Delta t)$ is the estimate of the stationary
connected correlation \eqref{eq:conn-corr-estimate}, then the integral
can be approximated by a discrete sum.  However, the sum cannot run
over all available values $k=0,\ldots,N-1$, because the variance of
$\CC_k$ for $k$ near $N-1$ is large (\S\ref{sec:comp-time-corr}), so
that the sum $\sum_{k=0}^{N-1} \CC_k$ is dominated by statistical
noise (more precisely, its variance does not go to zero for
$N\to\infty$).  A way around this difficulty is
\cite[\S3]{sokal_monte_1997} to cut-off the integral at a time
$t_c = c \Delta t$ such that $c\ll N$ but the correlation is already
small at $t=t_c$ (i.e.\ $t_c$ is a few times $\tau$).  Thus $\tauint$
is defined self-consistently as
\begin{equation}
\tauint =  \int_0^{\alpha \tauint} \rho(t) \, dt ,
\label{eq:tauauto}
\end{equation}
where $\alpha$ should be chosen larger than about $5$, and within a
region of values such that $\tauint(\alpha)$ is approximately
independent of $\alpha$.  Longer tails will require larger values of
$\alpha$; we have found adequate values to be as large as 20
\citep{Cavagna2012}.  To solve \eqref{eq:tauauto} one can compute
$\tau(M)=\sum_k^M \CC_k/\CC_0$ starting with $M=1$ and increasing $M$
until $\alpha\tau(M)>M$.

Another useful definition of correlation time can be obtained from
$\tilde\rho(\omega)$, the Fourier transform of $\rho(t)$.
Normalisation of $\rho(t)$ implies that
$1=\rho(0) = \int_{-\infty}^{\infty} \frac{d\omega}{2\pi}
\,\tilde\rho(\omega)$.  Then a characteristic frequency $\omega_0$
(and a characteristic time $\tau_0=1/\omega_0$) can be defined such
that half of the spectrum of $\tilde\rho(\omega)$ is contained in
$\omega\in[-\omega_0,\omega_0]$ \cite{halperin_scaling_1969}, i.e.\
\begin{equation}
 \int_{-\omega_0}^{\omega_0} \frac{d\omega}{2\pi} \,\tilde\rho(\omega)=\frac{1}{2}.
 \label{eq:HHomegak}
\end{equation}
This definition of can be expressed directly in the time domain writing
\begin{equation}
 \frac{1}{2} = \int_{-\omega_0}^{\omega_0} \frac{d\omega}{2\pi}
 \int_{-\infty}^\infty \!\!dt \, \rho(t) e^{i\omega t} =
 2 \int_0^\infty \!\!dt \rho(t) \int_{-\omega_0}^{\omega_0} \frac{d\omega}{2\pi}
   e^{i\omega t} = \frac{2}{\pi} \int_0^\infty \!\!dt \, \rho(t)
   \frac{\sin\omega_0 t}{t},
\end{equation}
where we have used the fact that $\rho(t)$ is even.  Then the
correlation time is defined by
\begin{equation}
  \label{eq:HHrelaxtime}
   \int_0^\infty \!\! \frac{dt}{t} \, \rho(t)
   \sin\left(\frac{t}{\tau_0}\right) = \frac{\pi}{4}.
\end{equation}
It can be seen that if $\rho(t) = f(t/\tau)$, then $\tau_0$ is
proportional to $\tau$ (it suffices to change the integration variable
to $u=t/\tau$ in the integral above).
An advantage of this definition is that it copes well with the case
when inertial effects are important and manifest in (damped)
oscillations of the correlation function (see Fig.~\ref{fig:shapes}).

\subsection{Correlation length}
\label{sec:correlation-length}

A quite general procedure is to use the \emph{second moment
  correlation length} $\xi_2$ \citep[\S 2.3.1]{amit2006},
\begin{equation}
  \xi_2 = \sqrt { \frac{\int \!\! d\rr\, r^2 C_c(\rr)}
    {\int\!\!d\rr\, C_c(\rr)} }.
  \label{eq:def-xi2}
\end{equation}
The quotient here ensures that $\xi_2$ scales with control parameters
like $\xi$ (a definition analogous to that of the integral correlation
time would not have this property due to the power-law prefactor).
$\xi_2$ can also be expressed in terms of the Fourier transform of
$C_c(\rr)$ as
\begin{equation}
  \xi_2^2 = - \left. \frac{1}{\tilde C(\kk)} \frac{\partial^2 
      \tilde C(\kk)}{\partial k_\alpha \partial k_\alpha}
  \right\rvert_{k=0}.
  \label{eq:xi2-fourier}
\end{equation}
The integrals in \eqref{eq:def-xi2} can be computed numerically over
the finite volume, but \eqref{eq:xi2-fourier} may be more convenient
(recall that $\hat C(\kk)$ can be computed directly from the data,
\S\ref{sec:estim-space-corr}).  Numerical derivatives should clearly
be avoided; instead one can fit $\tilde C(k)^{-1} = A + B k^2$ for
small $k$, and obtain the correlation length as $\xi_2 =\sqrt{ B/A}$
(which follows from \eqref{eq:xi2-fourier} apart from a
dimensional-dependent numerical prefactor).  Another possibility is to
obtain $A$ and $B$ using only $\tilde C(0)$ and
$\tilde C(k_\text{min})$ \citep{cooper_solving_1982}, where
$k_\text{min}$ is the smallest wavenumber allowed by boundary
conditions (e.g.\ $k_\text{min}=2\pi/L$ in a cubic periodic system),
resulting in
\begin{equation}
  \xi_2 = \frac{1}{k_\text{min}} \sqrt{
    \frac{\hat C(0)}{\hat C(k_\text{min})} -1}.  
\end{equation}
When working with lattice systems, $2\sin (k_\text{min}/2)$ can be
substituted for the $k_\text{min}$ outside the square root
\citep{caracciolo_wolff-type_1993}, rendering the expression exact
for a Gaussian lattice model.  This procedure works well when $\xi\ll
L$, so that the small wavevectors are seeing an exponential
behaviour for $r\sim L$.

If $A/B$ is not much larger than $L^{-2}$, then $\xi$ is approaching
the system size $L$ and the estimate is not very reliable; for $\xi$
of the order of $L$ or larger, $A/B$ may even become negative.  In
situations like these (which include critical systems), the first root
of $\CC_\text{sp}(r)$, can provide a useful proxy for the correlation scale, but it
is necessary to compare systems of different sizes
(see~\S\ref{sec:finite-size-effects}).


\section{Correlation functions in the critical region}
\label{sec:corr-funct-crit}

The critical region refers to the volume of parameter space near the
critical surface, which is a boundary between different phases of a
system.  The word phase may refer to a thermodynamic equilibrium
phase, but it also applies to systems out of equilibrium, where
different phases are characterised by qualitatively different
properties (static or dynamic).

The presence of long-range correlations is a key feature of critical
systems, so that correlation functions are a fundamental tool to study
systems at or near criticality.  Note, however, that long-range
correlations are not synonymous with the critical point: in systems
with a continuous symmetry (such as the classical Heisenberg model, in
which spins are vectors on the sphere), the correlation length is
infinite across all the broken-symmetry (low-temperature, magnetised)
phase \citep[Ch.~11]{goldenfeld_lectures_1992}.

Sufficiently close to the critical surface, the fact that the
correlation is much larger than microscopic lengths allows to write
the correlation functions for large enough distances in the form of
so-called \emph{scaling relations.}  These are functional relations
where the dependence on the microscopic parameters enters only through
a few control parameters, which are often recast in terms of the
correlation length.  The relations involve an unknown function, which
gives the shape of the correlation decay but is independent of the
control parameters, or depends on a subset of them.  We state first
the scaling relations for the static and dynamic correlation functions
as applicable to infinite systems, and afterwards consider the case of
finite systems.

Scaling relations as presented below apply and are well understood
within equilibrium thermodynamics, where the system's (near) scale
invariance can be exploited using the renormalization group to explain
how scaling relations, as well as subleading corrections, arise
\citep{amit2006, wilson1974, cardy1996, Binney1992, itzykson1989,
  lebellacQuantumStatisticalField1991}.  In out-of-equilibrium, and in
particular biological, systems our understanding is less firm,
although the scale invariance found near critical points can still be
exploited \citep[see e.g.][]{tauber2014}.  Scaling has been
successfully applied to biological systems (although in this case it
is typically essential to take into account the system's finite
size, as discussed in \S\ref{sec:finite-size-effects}), but situations
might be encountered where matters are more complicated than presented
here, especially when disorder or networks with complicated topology
are involved.

\subsection{Scaling}
\label{sec:scaling}

The connected correlation function for an infinite system near
criticality for the case of a single control parameter
can be written as
\begin{equation}
  C_c(r) = \frac{1}{r^{d-2+\eta}} f(r/\xi) =
   \frac{1}{\xi^{d-2+\eta}} \hat f(r/\xi)
    \label{eq:cr-scaling}
\end{equation}
where $f(x)$ is a function that tends to 0 faster than any power law
for $x\to\infty$, and the second equality is just a different way of
writing the same scaling law, using $\hat f(x) = x^{-d+2-\eta}f(x)$.
The exponent of the power-law prefactor is conventionally written in
this way because $\eta$ is typically a small correction to $d-2$,
which is the exponent obtained from dimensional analysis
\citep[Ch.~7]{goldenfeld_lectures_1992}.  The exponent $\eta$ is
called an \emph{anomalous dimension,} and is one of the set of
critical exponents that describe the singular behaviour of
thermodynamic quantities near criticality \citep[see e.g.][]{Binney1992}

As stated, the scaling law \eqref{eq:cr-scaling} applies for large $r$
and finite $\xi$.  At short distances, non-universal dependence on
microscopic details may be observed, in particular note that the
correlation must be finite for $r=0$ Precisely at the critical point,
the correlation does not decay exponentially but as a power
law,\begin{equation} C(r)\sim r^{2-d-\eta}.
\end{equation}
In other words, for $\xi=\infty$ the scaling function becomes a
constant, even if it is not true that $f(0)$ is finite.  This decay is
scale-free, in the sense that if one chooses an arbitrary scale $r_0$,
$C(r)/C(r_0)$ is a function of $r/r_0$, i.e.\ it can be scaled with
the externally chosen length scale.  In contrast, if the decay is e.g.\
exponential, $C(r)/C(r_0)$ will depend on $r/r_0$ \emph{and} on
$r_0/\xi$, i.e.\ the function has an intrinsic length scale $\xi$.

If there is more than one control variable that can take the system away
from the critical surface then the scaling law can be more complicated
than~\eqref{eq:cr-scaling} \citep[see][Ch.~9]{goldenfeld_lectures_1992}.

\begin{figure}
  \includegraphics[width=\columnwidth]{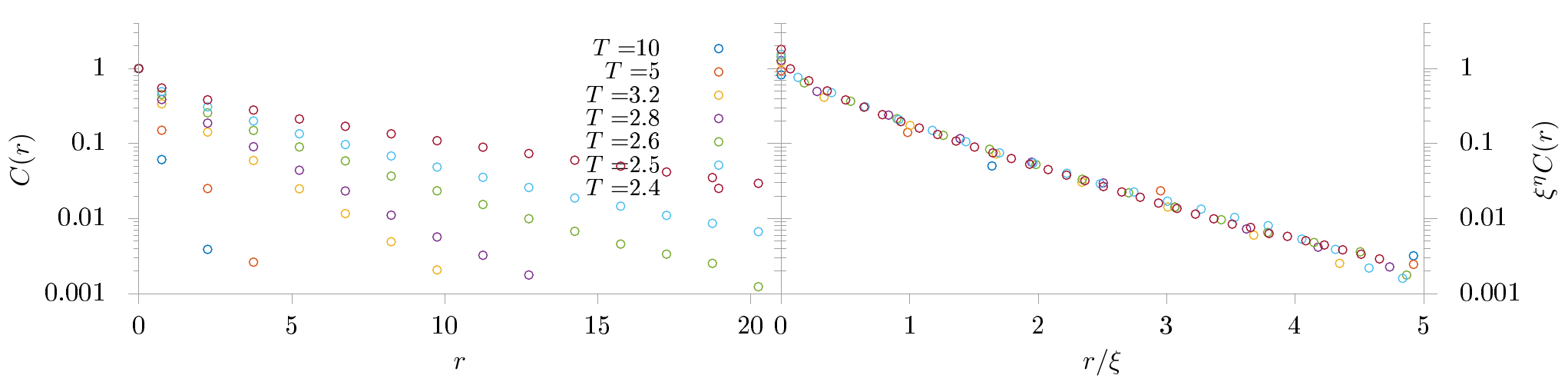}\\\includegraphics[width=\columnwidth]{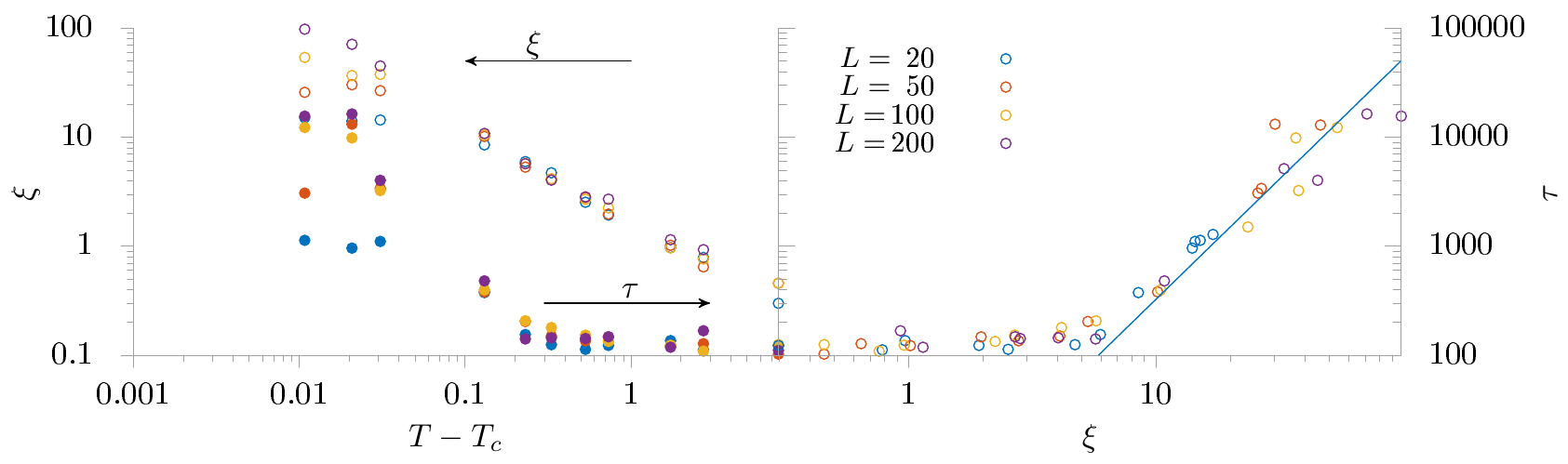}
  \caption{Example of static (top) and dynamic (bottom) scaling in the
    2-$d$ Ising model.  In the top panel, the space correlation
    function $C(r)$ was computed using
    eq.~\eqref{eq:space-cc-estimate} for the Isinng model on a
    $100\times100$ square lattice at the specified temperatures.  The
    correlation length $\xi_2$ was computed in $k$-space,
    Eq.~\eqref{eq:xi2-fourier} fitting the first five points of
    $C^{-1}(k)$ to a quadratic function.  The rightmost plot shows
    that $\xi^\eta C(r)$ plotted vs.\ $r/\xi$ falls on the same curve
    for all temperatures.  We have used the exact known value
    $\eta=1/4$.  Bottom panel: the correlation length and time
    (computed with definition \eqref{eq:HHrelaxtime}) for several
    temperatures and system sizes grow on approaching $T_c$, but
    saturate at a size-dependent value (left).  Right: correlation
    time vs.\ correlation length for the same sizes and temperatures
    as in the left panel, illustrating the dynamic scaling
    \eqref{eq:tau-vs-xi}. The is a power law with exponent $z=2.21$,
    value of the dynamic exponent obtained  by \citet{munkel1993}.}
  \label{fig:Ising-scaling}
\end{figure}

The scaling of $C_c(r)$ is illustrated in Fig.~\ref{fig:Ising-scaling}
for the 2-$d$ Ising model.

From \eqref{eq:cr-scaling} an equivalent scaling law can be written
for the correlaiton in Fourier space:
\begin{align}
  \tilde C(k) & = \xi^{2-\eta} F(k\xi) = k^{-2+\eta} \hat F(k\xi),
                \label{eq:ck-scaling}
\end{align}
where $\hat F(x) = x^{2-\eta} F(x)$ and now the scaling law is valid
for small $k$, with $F(x=0)$ finite, so that $\hat C(k=0)$ (which is
proportional to the susceptibility) diverges when $\xi\to\infty$.

The space-time correlation can also be written in a scaling form,
which ultimately leads to a link between $\xi$ and $\tau$.  This
\emph{dynamic scaling} relation states that
\citep{halperin_scaling_1969, hohenberg_theory_1977, tauber2014}
\begin{align}
  \tilde C(k,\omega) &= \tilde C(k) \frac{2\pi}{\omega_k}
                       h(\omega/\omega_k, k\xi),
                       & \omega_k = k^z \Omega(k\xi)
  \label{eq:dynscaling}
\end{align}
where $\tilde C(k)$ is the static correlation function and $\omega_k$
is the $k$-dependent inverse characteristic time defined
through~\eqref{eq:HHomegak} applied to $\tilde C(k,\omega)$ at fixed
$k$. Eq.~\eqref{eq:dynscaling} introduces the \emph{dynamic exponent}
$z$ and two scaling functions: $\Omega(x)$, which stays finite for
$x\to\infty$, and the shape function $h(x)$, which obeys
$\int_{-\infty}^\infty h(x)\,dx=1$ and $\int_{-1}^1 h(x)\,dx=1/2$
because
$C(k)=C(k,t=0)=\int_{-\infty}^\infty C(k,\omega)\,d\omega/2\pi$.  In
the time domain this scaling reads
\begin{align}
  C(k,t) &= C(k) \hat h(t/\tau_k,k\xi), & \tau_k = k^{-z}\Omega^{-1}(k\xi).
  \label{eq:dynscaling-time}
\end{align}
In general the correlation time \emph{and} the shape of the normalised
$k$-dependent time correlation depend on $k$.  If one compares systems
with different correlation lengths but at different $k$ such that the
product $k\xi$ is held constant, then $C(k,t)/C(k)$ will be seen to
depend on $t$ and $k$ only through the scaling variable $k^zt$.

One can multiply and divide by $\xi^z$ to rewrite the scaling of $\tau_k$:
\begin{equation}
  \tau_k = k^{-z} \Omega^{-1}(k\xi) =  \xi^z \hat \Omega^{-1}(k\xi),
  \label{eq:tauscaling}
\end{equation}
where $\hat\Omega^{-1}(x)=x\Omega^{-1}(x)$.
From~\eqref{eq:tauscaling} we learn the important fact that, at fixed
$\xi$, the correlation time depends on the observation scale, more
precisely that it is smaller for shorter length scales (larger $k$).
Since $\tau$ must be finite for finite $\xi$, $\hat\Omega^{-1}(x)$
must be finite for $x\to0$, and from the second equality we conclude
that the relaxation time of global quantities (i.e.\ at $k=0$) grows
with growing correlation length as
\begin{equation}
  \tau \sim \xi^z.  \label{eq:tau-vs-xi}
\end{equation}
This important relationship highlights the fact that static, or
equal-time, correlations, are nevertheless the result of a dynamic
process of information exchange, and that the farther this information
travels, the slower the system becomes.  So static correlations should
not be viewed as instantaneous correlations.  Indeed they contain
contributions from all frequencies, since
$C(k,t=0)=\int_{-\infty}^\infty \tilde C(k,\omega) \, d\omega/(2\pi)$.

\subsection{Finite-size effects}
\label{sec:finite-size-effects}

Scaling laws as written in the previous subsection apply only to
infinite systems.  Although infinite systems exist only in theory,
experimental systems can be considered infinite when $\xi$ is much
smaller than the system's linear size $L$.  It is true that when the
critical point is approached, $\xi$ will eventually grow to be larger
than any system size.  However, in condensed matter systems samples
are typically many orders of magnitude larger than microscopic
lengths, and $\xi$ can be extremely large with respect to
inter-particle distances while still being smaller than $L$.  The
condition $\xi\ge L$ will only be true for a range of temperatures so
thin around the critical point that it this effect can in practice be
ignored.

In contrast, in experiments in biology the number of units making up
the system (aminoacids, cells, organisms) is quite far from the
$10^{23}$ or so that condensed matter systems can boast, and the
effects of finite size must be properly taken into account.  The same
is true for numerical simulations, where in fact finite-size effects
have long been used \citep{privman1990} to extract information about
the critical region.  More recently, has also been exploited in
experiments on biological systems.

Finite-size effects blur the sharp transitions and singularities that
distinguish phase transitions in the thermodynamics of infinite
systems.  If a critical system is finite, it can be scale invariant
only up to the its size $L$; in this sense $L$ acts as an effective
correlation length if $\xi>L$.  Conversely, if $\xi$ is finite but
larger than $L$, the system will appear nevertheless scale invariant.
Thus $L$ is finite, scaling relations must be modified to account for
the fact that correlations cannot extend beyond $L$.  These modified
relations are known as \emph{finite-size scaling} relations
\citep{barber1983, cardy1988}.

According to the finite-size scaling \textsl{ansatz}
\citep{barber1983}, when $\xi$ is of the order or greater than $L$ the
scaling relations must be modified to include the ratio $L/\xi$.  This
\textsl{ansatz} can be justified with renormalisation group
argouments \citep{amit2006, cardy1988}.  For the static
correlation function, instead of~\eqref{eq:cr-scaling} we write
\begin{align}
  C(r) &= \frac{1}{r^{d-2+\eta}} f(r/\xi,L/\xi),  & \text{if } L>\xi,
\end{align}
with $f(x,y)\to f(x)$ for $y\to\infty$, and
\begin{align}
  C(r) &= \frac{1}{r^{d-2+\eta}} g(r/L,L/\xi)  & \text{if } L<\xi,
\end{align}
with $g(x)$ another scaling function.


At the critical point, we are effectively left
with a scaling relation of the same type as \eqref{eq:cr-scaling} but
with $L$ replacing $\xi$ \cite{cardy1988},
\begin{equation}
  C(r) = \frac{1}{r^{d-2+\eta}} g(r/L) 
  = \frac{1}{L^{d-2+\eta}} \hat g(r/L) 
\end{equation}
(and a different scaling function).  Correlations are truly scale-free
only in the limit $L\to\infty$, where the scaling function is replaced
by a constant.  In a finite system, function $g(x)$ will introduce a
modulation of the power law when $r\sim L$.  Scale invariance
manifests instead as a correlation scale that is proportional to $L$:
enlarging the system also enlarges the correlation scale.  This,
together with the fact that the space-averaged correlation function
has at least one zero, can be exploited to show that a system is
scale-invariant.  Calling $r_0$ the first zero of $C_\text{sp}(r)$,
one can show \citep[\S2.3.3]{cavagna_physics_2018} that
\begin{subequations}
  \label{eq:r0FS}
  \begin{align}
    r_0 &\sim \xi \log(L/\xi), & L\gg\xi, \\
    r_0 & \sim L, & L\ll \xi.   \label{eq:scale-free-fs}
  \end{align}
\end{subequations}
We emphasise that $r_0$ is \emph{not} in general a correlation length,
since it behaves differently: it does not have a finite limit for
$L\to\infty$ (even away from the critical point) and it does not scale
like $\xi$ with control parameters.  However, \emph{if} $\xi=\infty$,
then $r_0$ is proportional to $L$ (the only correlation scale).  So,
if one can show that~\eqref{eq:scale-free-fs} holds for a set of
systems of different sizes, but otherwise equivalent, one can argue
that those systems are in fact scale-free.  Such reasoning has been
used e.g.\ in \citet{tang_critical_2017, Cavagna2010, Attanasi2014,
  fraiman_what_2012}.

In Fourier space, the finite-size version of \eqref{eq:ck-scaling} can
be written, for $L<\xi$,
\begin{equation}
  \label{eq:5}
  \hat C(k;L) = k^{-2+\eta}F(kL,L/\xi) = L^{2-\eta} \hat F(kL,L/\xi).
\end{equation}

The finite-size version of the dynamic scaling law
\eqref{eq:dynscaling-time} is
\begin{equation}
  \label{eq:7}
  \tilde C(k,t;L) = \tilde C(k;L) \hat h(t/\tau_k,kL,L/\xi), \qquad \tau_k=k^{-z}\Omega^{-1}(kL,L/\xi)
\end{equation}
Then, at the scale-free point $\xi=\infty$ it holds that
$\tilde C(k,t;L)$ measured for systems of different size depends on $k^zt$ if one
measures each system at a $k$ such that $kL=\text{const}$.  This
scaling law has been successfully applied to study the dynamical
behaviour of swarms of midges in the field
\citep{cavagna_dynamic_2017}.  For the global relaxation time one has,
instead of \eqref{eq:tau-vs-xi}
\begin{equation}
  \tau = L^z \hat\Omega^{-1}(L/\xi).
\end{equation}

To conclude, let us point out that from what we have said it follows
that a single measurement of $\xi$ on a finite system does not tell
one anything about whether correlations are long range or not.  An
unknown numerical prefactor is involved in the estimation of $\xi$,
and the estimators we have discussed tend to be less reliable when
$\xi\sim L$, so that it does not make sense to discuss whether, say,
is $\xi=L/2$ large or not.  Instead, it is necessary to measure the
trend of $\xi$ with control parameters, and if at all possible, with
$L$.  Moreover, studying different system sizes can show whether the
correlation scale grows with $L$ and establish that the system is
scale-free in situations where changing the control parameters is not
feasible.

There can be situations, especially in biological experiments, where
changing system size is not possible (e.g.\ how would one study human
brains of different sizes?).  In such circumstances, one may attempt
to do finite-size scaling by measuring subsystems (``boxes'') of size
$W\ll L$ and studying how the quantities change when varying $W$.  The
idea can be traced back to \citet{binder1981}, and has also been
employed out of equilibrium \citep{fernandez2015}.  The applicability
of relations \eqref{eq:r0FS} with the box size $W$ in place of $L$ for
systems of fixed size was recently demonstrated for several simulated
systems \citep{martin2020a}.

\section{Space-time correlations in neuronal networks}
\label{sec:dynam-scal-neur}

The finding of scale-free avalanches in cortical tissue
\citep{beggs2003, shew2009} and strong correlations in retinal cells
\citep{schneidman2006, tkacik_ising_2006, tkacik2015} are early
experimental evidence of criticality in neuronal populations (see
\citet{Mora2011} and \citet{chialvo_emergent_2010} for reviews of
these and more recent experiments). At the level of the whole brain,
fMRI measurements of neuronal activity \citep{expert2011,
  fraiman_what_2012, tagliazucchi_criticality_2012} have found
long-range correlation, providing support for the conjecture that the
brain operates at a critical point \citep{bak1996, chialvo1999,
  beggs2008}.

Notwithstanding the importance of these works in forwarding our
understanding of how complex behaviour can arise in the brain, it is
fair to say that the picture of the brain as a critical system is
still work in progress, and surely new experiments will be attempted
in the near future.  Critical systems display distinctive dynamic, as
well as static characteristics, and much information can potentially
be gained by studying systematically static and dynamic correlations
together on the same system, and in particular the scaling relations.
For instance, study of static correlations in midge mating swarms
using finite-size scaling led to the finding \citep{Attanasi2014} that
they are in a critical regime compatible with the critical Vicsek
model \citep{Vicsek1995a}.  A subsequent investigation of dynamic
correlations in the same system showed that they obey dynamic scaling
\citep{cavagna_dynamic_2017}, but with a novel dynamic exponent,
incompatible with the dynamics of the Vicsek model.  The exponent
turned out to be different from those of the classical dynamic models
\citep{hohenberg_theory_1977}, and this spurred still-ongoing efforts
to understand whether a combination of inertia and activity
ingredients can explain this new exponent \citep{cavagna2019a}.

Clearly, magnets, swarms and brains are quite different systems, and
in particular the special network structure of neuronal assemblies
poses specific problems \citep{korchinski2021}.  Certainly it is not
the case to take tools developed in the statistical mechanics of
condensed matter and blindly apply them to study the brain.  But
knowledge of other critical systems suggests the relation between
correlation time and length scales can hold valuable information.  The
question calls for an experimental study, but we wish to conclude this
article with an exploration of a simple excitable network model, which
can serve as some illustration to the exposition of correlation
functions, and hopefully stir up interest for experimental studies of
length and time correlation scales in the brain.

\subsection{The ZBPCC model}

We study the simple neural network model proposed by
\citet{zarepour2019a}, which is essentially the model used by
\citet{haimovici_brain_2013} but implemented on a small-world network.
The dynamics is that of a Greenberg-Hastings cellular automaton
\cite{greenberg1978}, where each site of the network has three
possible states, $S_i=\{Q,E,R\}$, corresponding to quiescent, excited
or refractory, respectively, and the transitions among them are ruled
by the probabilities
\begin{subequations}
\begin{align}
  P_{i,Q\to E} &= 1 - \bigl[1-r_1\bigr]\left[ 1-  \Theta \left( \sum_j W_{ij} \delta_{S_j,E} -T \right)  \right], \label{eq:act-modA} \\
  P_{i,E\to R} & =1, \\
  P_{i,R\to Q} & = r_2.
\end{align}
\end{subequations}
$P_{i,a\to b}$ is the probability that site $i$ will transition from
$a$ to state $b$, $\Theta(x)$ is Heaviside's step function
[$\Theta(x)=1$ for $x\ge0$ or 0 otherwise], $\delta_{i,j}$ is
Kronecker's delta, $r_1$, $r_2$ and $T$ are numeric parameters and
$W_{ij}$ is the network's connectivity matrix (see below).  Thus an
active site always turns refractory in the next time step, and a
refractory site becomes quiescent with probability $r_2$.  We set
$r_2=0.3$ as in previous work \citep{zarepour2019a,
  haimovici_brain_2013}, so that the site remains refractory for a few
time steps, and $r_1=10^{-6}$ so that external excitation events are
relatively rare.  The probability for a quiescent site to become
active is written as 1 minus the product of the probabilities of
\emph{not} becoming active through the two mechanisms at work:
\emph{spontaneous activation,} which occurs with a small probability
$r_1$ (mimicking an external stimulus), or \emph{transmitted
  activation,} which occurs with certainty if the sum of the weights
of the links connecting $i$ to its active (excited) neighbours exceeds
a threshold $T$.  All sites are updated simultaneously.

The model runs on an undirected weighted graph described by the
connectivity matrix elements $W_{ij}\ge0$, where a nonzero element
means a bond with the specified weight connects sites $i$ and $j$, and
$W_{ij}=W_{ji}$.  Neither the connectivity nor the weights depend on
time.  The graph is a bidirectional Watts-Strogatz small-world network
\citep{watts1998} with average connectivity $\langle k \rangle$ and
rewiring probability $\pi$.  The network is constructed as usual
\citep{watts1998} by starting from ring of $N$ nodes, each connected
symmetrically to its $\langle k\rangle /2$ nearest neighbours; then
each link connecting a node to a clockwise neighbour is rewired to a
random node with probability $\pi$, so that average connectivity is
preserved.  The rewiring probability is a measure of the disorder in
the network; $\pi=0$ preserves the original network topology, while
$\pi=1$ gives a completely random graph.  Once the network bonds are
drawn, each is assigned a random weight drawn from an exponential
distribution, $p(W_{ij}=w)=\lambda e^{-\lambda w}$, with
$\lambda=12.5$ chosen to mimic the weight distribution of the human
connectome \citep{zarepour2019a}

The simplest way to characterise the state of the network is perhaps
the activity order parameter, i.e.\ the fraction of active sites
\begin{equation}
  \label{eq:8}
  a = \frac{1}{N} A, \qquad A=\sum_i \delta_{S_i,E}.
\end{equation}
By analogy to magnetic systems, one can define an associated
susceptibility, proportional to the variance of $a$,
\begin{equation}
  \label{eq:9}
  \chi = \frac{1}{N} \left[ \langle A^2 \rangle - \langle A\rangle^2\right],
\end{equation}
where $\langle\ldots\rangle$ is a time average.  Another possible
characterisation is by studying the percolation of active clusters, as
in \citet{zarepour2019a}.  The percolation order parameter is the
probability that a site belongs to the largest cluster,
\begin{equation}
  P_\infty = \frac{1}{N} S_1,
\end{equation}
where $S_1$ is size of the largest active cluster, i.e.\ the number of
sites that belong to the largest connected component of the graph of
active sites.  The quantity that plays a role analogous to $\chi$ is
the mean (finite) cluster size \cite{stauffer1994},
\begin{equation}
  S = \frac{ \sum'_s s^2 N_s } {\sum'_s s N_s },
\end{equation}
where $N_s$ is the number of clusters of size $s$, the sum is over all
possible values of $s$ and the prime means that the largest cluster is
excluded.

\begin{figure}
  \includegraphics[]{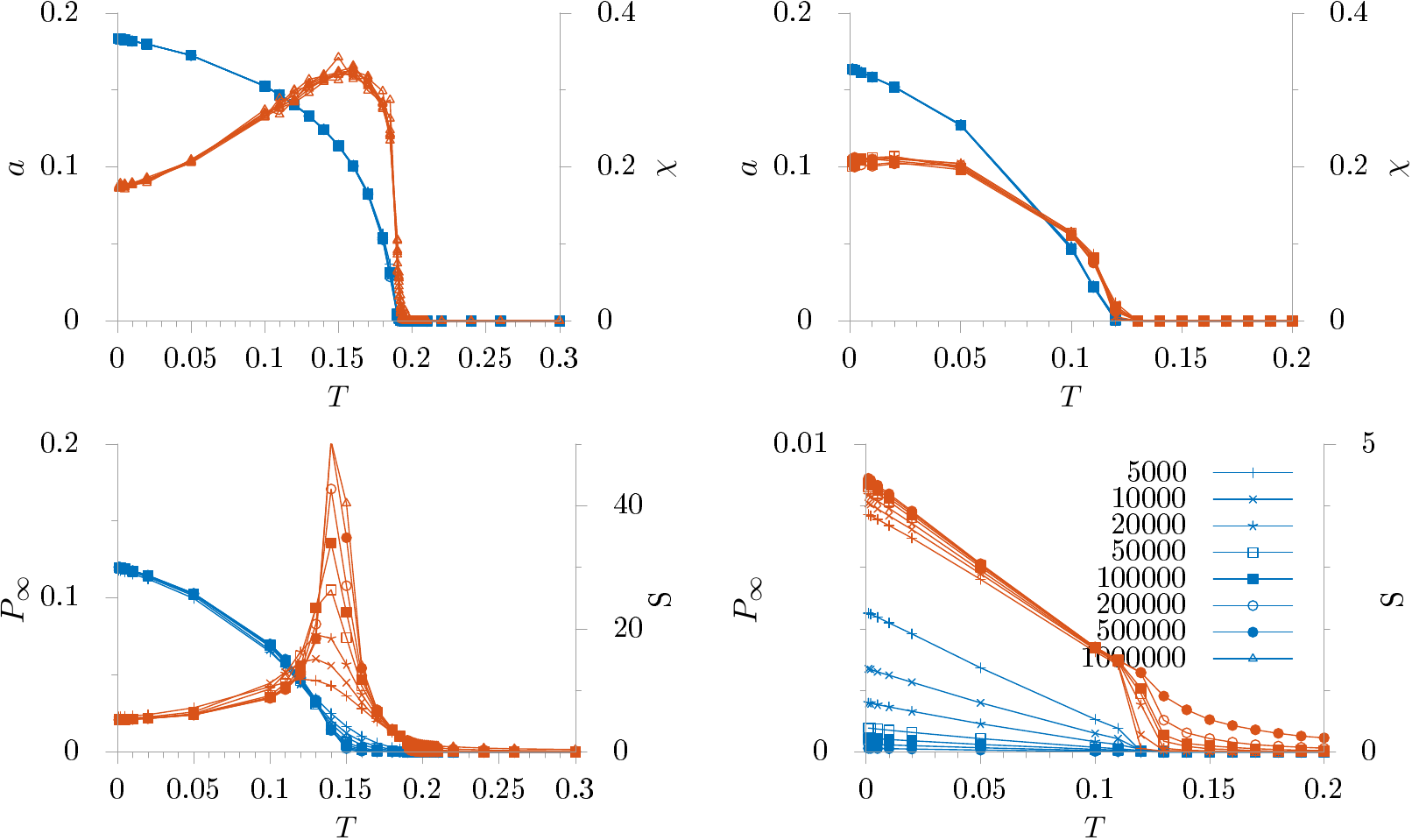}
  \caption{Dynamic transition of the ZBPCC model on the Watts-Strogatz
    small world network with average connectivity $\mean{K}=12$ (left
    column) and $\mean{K}=6$ (right column).  Top: activity (blue) and
    susceptibility (orange) vs.\ threshold level $T$.  Bottom:
    percolation order parameter $P_\infty=S_1/N$ (blue) and mean
    cluster size $S$ (orange) vs $T$.  Symbol type indicates the
    network size.  Curves are obtained after averaging over 2 to 10
    realisations of the network. Sizes are $N=5\cdot 10^3$, $10^4$,
    $2\cdot 10^4$, $5\cdot 10^4$, $10^5$, $2\cdot 10^5$, $5\cdot 10^5$
    and $10^6$.  In all cases $\pi=0.2$, $r_1=10^{6}$, $r_2=0.3$.}
  \label{fig:zarepour-static}
\end{figure}

These quantities were computed as a function of $T$ for $\pi=0.2$ and
average connectivity $\mean{K}=12$ and $\mean{K}=6$
(Fig.~\ref{fig:zarepour-static}).  At this value of $\pi$,
$\mean{K}=6$ lies in the ``no transition'' region, while $\mean{K}=12$
is in the continuous transition region of $(\pi,\mean{K})$ space
according to Fig.~4 of \citet{zarepour2019a}.  Indeed we find $\mean{K}=12$
shows a clear percolation transition (Fig.~\ref{fig:zarepour-static}
bottom left), with a peak in $S$ that grows with system size and
$P_\infty$ that goes to zero more and more sharply at
$T_p\approx 0.15$.  Instead for $\mean{K}=6$ no transition is observed,
and $P_\infty$ becomes progressively flatter as $N$ is increased
(Fig.~\ref{fig:zarepour-static} bottom right).

However, study of percolation requires knowledge of network
connectivity, which may be difficult to obtain experimentally.  A
simpler approach is to measure the average activity, without regards
to connectivity.  Using $a$ as order parameter one obtains a different
picture: both cases appear to have a continuous transition, where
activity goes smoothly to zero at $T_a\approx 0.19$ for $\mean{K}=12$
and at $T_a\approx 0.12$ for $\mean{K}=6$.  The transition is however
unusual in that $\chi$ has a (not very sharp) maximum at a different
value of $T$, and the height of the peak in $\chi$ does not scale with
size (i.e.\ unlike usual second order transitions, here $\chi$ never
diverges).  We have found no sign of the two transition thresholds
$T_p$ and $T_a$ moving toward each other, up to our maximum network
size of $10^6$ nodes.

\begin{figure}
  \includegraphics[]{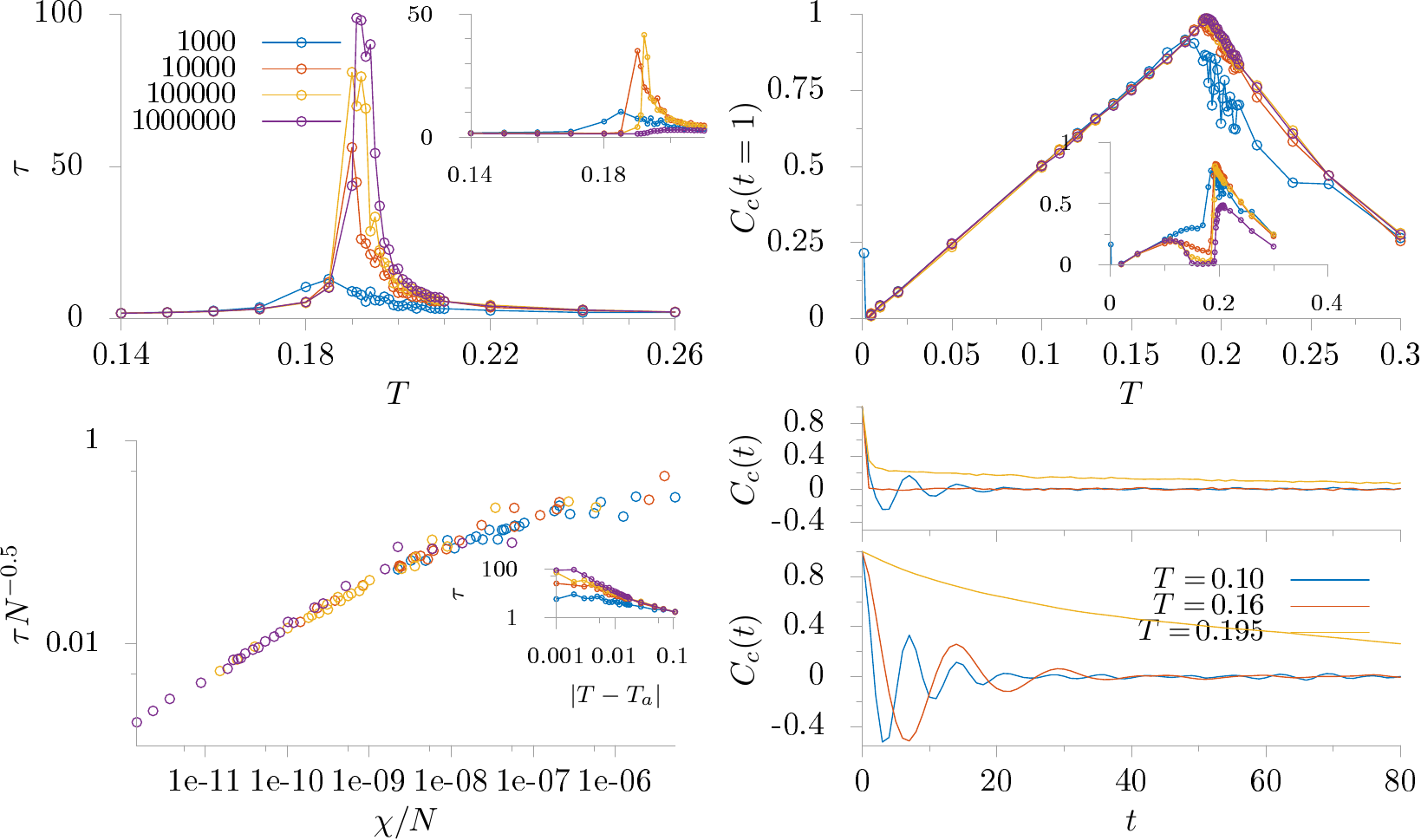}
  \caption{Correlation time and correlation function on the ZBPCC
    model for $\mean{K}=12$ (corresponding to the right column of
    Fig.~\ref{fig:zarepour-static}).  The relaxation time of the
    activity order parameter has a peak at the activity transition
    $T_a$ (top left), while no peak is observed near $T_P$, even in
    the correlation time of the percolation order parameter $S_1/N$
    (top left inset).  Symbol colour indicates system size.
    $C_c(t=1)$ also has a peak at $T_a$ (top right) when computed for
    the activity, while for $S_1/N$ two small peaks seem to develop at
    large $N$ (top right).  The maximum correlation time $\tau_a$
    grows with system size (bottom left inset) and can be scaled
    against $\chi/N$ (bottom left).  The time correlation function of
    the activity is damped oscillatory below $T_a$ but does not
    oscillate above $T_a$ (lower plot, bottom right), and the same
    for the correlation of $S_1/N$ above and below $T_p$ (upper plot,
    bottom right).}
  \label{fig:zarepour-dynamics}
\end{figure}

Which of the two transitions should we pay attention to?  In a sense
both are complementary, since they reveal different aspects of the
network activity.  However, a study of time correlations shows that
the activity transition, though less well-behaved from the static
point of view, is more relevant dynamically.  We have computed the
time correlation function of both order parameters ($A/N$ and
$S_1/N$), and the corresponding correlation times $\tau_a$ and
$\tau_s$ defined according to \eqref{eq:HHrelaxtime} (see
Fig.~\ref{fig:zarepour-dynamics}).  The correlation time of the
activity, $\tau_a$, shows a peak that grows with system size at the
activity transition $T_a$, while the correlation time $\tau_s$ has a
broader peak also near $T_a$, but it does not grow with $N$.  A
similar situation is found looking at the normalised time correlation
function after one step, $C_c(t=1)$, a quantity which peaks at a
regular second order phase transition \citep{chialvo2020}: when
computed for the activity, it displays a clear peak at $T_a$, while
when computed for the largest cluster size it shows, for large
systems, two small peaks, near $T_a$ and near $T_P$, although it is
not clear whether these will survive the thermodynamic limit.

At both transitions, the shape of the time correlation of the
respective order parameter changes, with damped oscillations found at
low $T$ giving way to an overdamped-like decay at higher $T$ (lower
right panel of Fig.~\ref{fig:zarepour-dynamics}).

The plot of $\tau_a$ vs.\ $T$ shows a peak that grows with system
size, with curves of $\tau_a$ vs.\ $\lvert T-T_a\rvert$ saturating at
higher values for larger systems.  Unfortunately, it is not possible
to explore standard dynamic scaling in this model, because it is
defined on a graph with no spatial structure, so that space
correlations cannot be defined.  We make an attempt using $\chi$ (which in
a ferromagnetic transition would scales with a power of $\xi$,
although here it never diverges).  Somewhat surprisingly, a
reasonable scaling behaviour is obtained plotting $N^{-1/2}\tau$ vs
$\chi/N$, i.e.\ the variance of $a$.

\subsection{ZBPCC on the fcc lattice}

We explore the simplest strategy to endow the ZBPCC model
with a spatial structure: we build a graph connecting the nearest
neighbours of a regular lattice, and then rewire it with probability
$\pi$ as in the small world case.  We chose the face-centred cubic
(fcc) lattice, which has connectivity $K=12$, the same value where for
the small-world case we found both activity and percolation transitions.

\begin{figure}
  \includegraphics[]{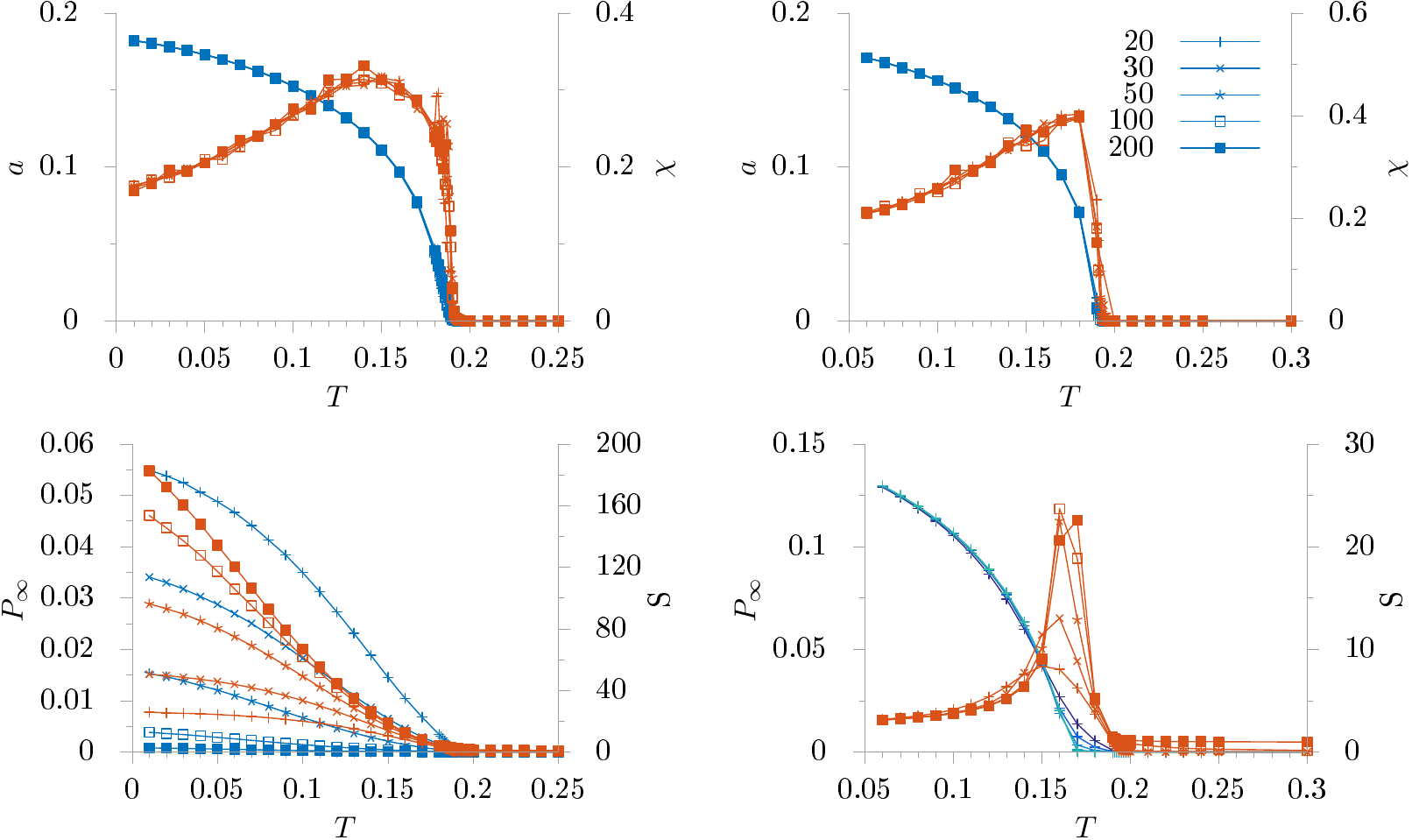}
  \caption{ZBPCC on the fcc lattice (left column) and on the fcc
    lattice rewired with $\pi=0.2$ (right column). Top row: activity $a$ (blue)
    and susceptibility $\chi$ (orange); bottom row: $P_\infty$ (blue)
    and $S$ (orange).}
  \label{fig:ZBPCC-fcc-static}
\end{figure}

Fig.~\ref{fig:ZBPCC-fcc-static} shows the behaviour of the activity
and percolation order parameters for the pure fcc lattice ($\pi=0$)
and for the fcc lattice rewired with $\pi=0.2$.  The fcc lattice shows
a situation similar to the small world network with $\mean{K}=6$, with
a transition in $a$ but no percolation transition.  The maximum of
$\chi$ is shifted with respect to $T_a$.  The rewired fcc lattice
instead shows both the activity and percolation transitions, in this
case with the maximum of $\chi$ very close to $T_a$.  Here
$T_P\approx 0.17$.  In both cases $T_a\approx 0.19$.  These results
suggest that $T_a$ depends mainly on the network connectivity, while
the percolation transition is more sensitive to the details of network
topology.

Note that the threshold for random site percolation on
the fcc lattice place the percolation transition at
has been estimated at an occupation probability of
$p_c\approx 0.199$ \citep{vandermarck1998, xu2014}.  The recorded
network activity is always below $0.190$, so that, if active sites
were distributed at random, one would not expect a percolating cluster
to form.  The fact that an infinite cluster appears for $\pi=0.2$ at
about $a\approx0.094$ means that active sites must be more clustered
than would result from random activation.

\begin{figure}
  \includegraphics[]{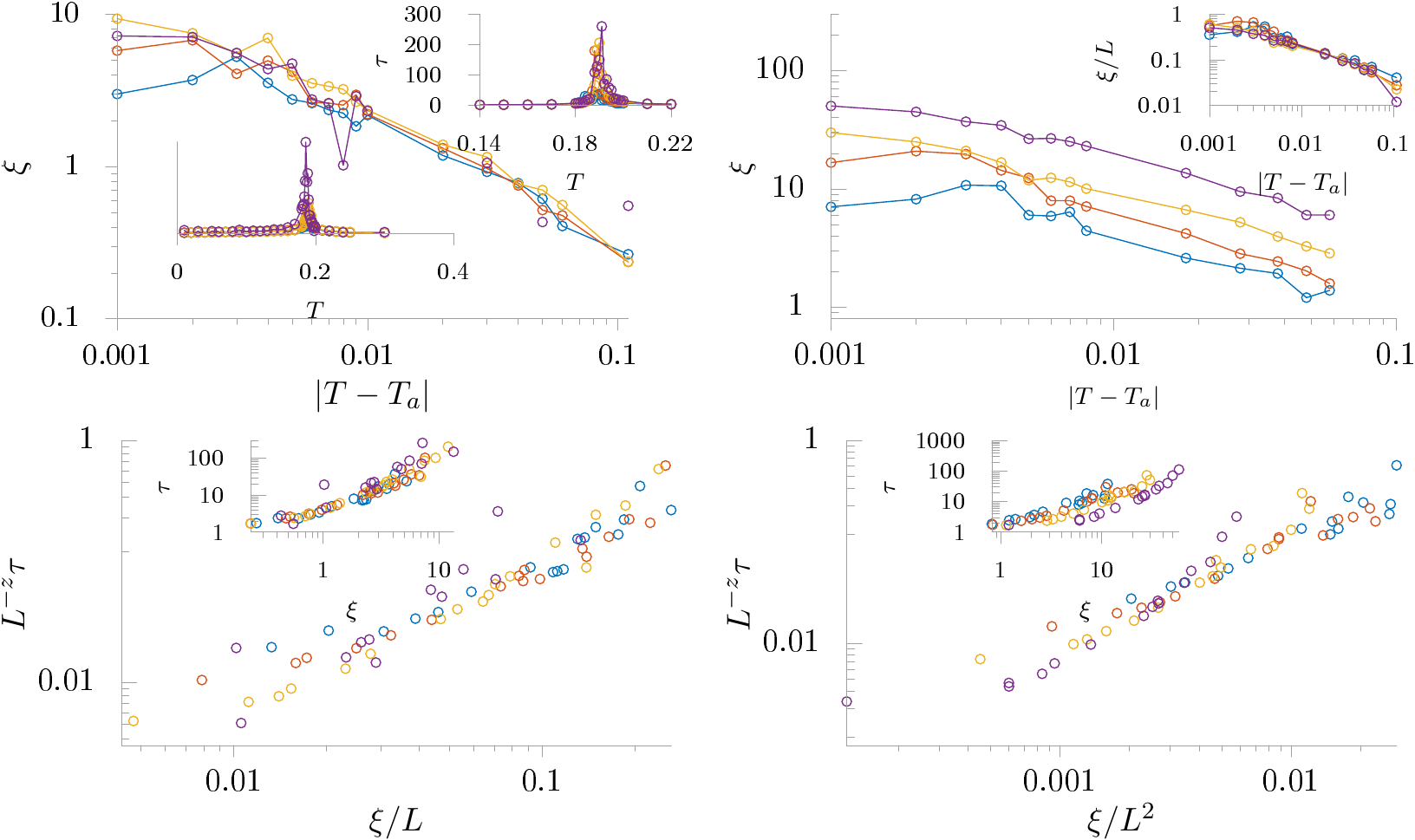}
  \caption{ZBPCC on the fcc lattice (left column) and the fcc rewired
    with $\pi=0.2$ (right column).  The correlation length and time
    have a peak at $T_a$ (top left). In the case of the rewired
    network, $\xi$ has a strong dependence on lattice size, indicated
    by symbol color (top right).  Bottom row: relationship between
    $\tau$ and $\xi$.  }
  \label{fig:ZBPCC-fcc-dynamic}
\end{figure}

In this version of the model we can use the underlying fcc lattice to
define distances and compute space correlations.  We have computed
$C(k)$ and used it to compute $\xi_2$ with definition
\eqref{eq:xi2-fourier} and a quadratic fit at small $k$ (see
Fig.~\ref{fig:ZBPCC-fcc-dynamic}).  For $\pi=0$ the largest $\xi$ is
about a tenth of the lattice size, and finite-size effects are modest.
On the contrary, on the rewired fcc, there is a strong size effect.
However, instead of the situation usual in a regular lattice, with
$\xi$ vs. $T$ curves superimposing at high $T$ and saturating at
different values on approaching the critical point (see
Fig.~\ref{fig:Ising-scaling}), here $\xi$ grows with $L$ almost
uniformly for all $T$, and the $\xi$ vs $T$ seem to scale when
dividing by $L$ (Fig.~\ref{fig:ZBPCC-fcc-dynamic}, inset of top right
panel).  Given that this effect only appears with rewiring, it may be
related to the fact that when a link is rewired, the new neighbour is
chosen randomly over all the nodes without regards to the distance, a
procedure that is not very realistic.

Finally we have computed the correlation time of the activity.  This
shows a size-dependent maximum at $T_a$ in both cases, confirming the
dynamical relevance of $T_a$.  Comparing $\tau_a$ and $\xi$ shows that
both grow together, but the attempt at applying dynamic scaling
(bottom panels of Fig.~\ref{fig:ZBPCC-fcc-dynamic}) is not very
satisfactory.  A more detail analysis is needed of the role played by
topological details and network disorder.

\section{Conclusions}
\label{sec:conclusions}

We have discussed how space (static), time (dynamic), and space-time
correlation functions can be defined and computed in a variety of
situations, how characteristic correlation length and times are
computed, and summarised our knowledge of scaling relations and
finite-size effects for systems near a critical point.  We have
finally applied these tools to analyse the dynamic transition of the
simple ZBPCC neuronal model, as well as a variant thereof introducing
a simple spatial structure.  Studying the time correlations of the two
order parameters considered showed that only the activity correlation
time has a peak, thus hinting that the activity transition is more
relevant for the description of the model dynamics.  The fcc version
of the ZBPCC proposed here showed a correlation length that also peaks
at the activity transition, but with an unusual finite-size behaviour,
especially in the rewired case.  The way space has been introduced in
the model is however somewhat artificial, and it is worthwhile to
consider models with realistic spatial structure and connectivity,
implementing a simple excitable model like the Greenberg-Hastings but
on the human connectome, as in \citet{haimovici_brain_2013} or
\citet{odor2016}.  Work in this direction is in progress.

I thank D.~Chialvo for helpful discussions on the ZBPCC model, and for
bringing some references to my attention.  I also thank A.~Cavagna and
I.~Giardina, who over a long collaboration helped shape my
understanding of concepts and practicalities of correlation functions,
especially as applied to biological systems.  This work was supported
in part by grants from Universidad Nacional de La Plata and Consejo
Nacional de Investigaciones Cient\'\i{}ficas y T\'ecnicas (CONICET,
Argentina).

\appendix


\section{Stochastic processes}
\label{sec:stochastic-processes}

A \emph{stochastic process}, or sometimes \emph{random field,} is a
family of random variables indexed by $\rr$ and $t$.  Thus there must
exist a family of probability densities $P(a,\rr,t)$ that allows to
compute all moments $\mean{a^n(\rr,t)}$ and in general any average
$\mean{f\bigl(a(\rr,t)\bigr)}$.  But $P(a,\rr,t)$ is not enough to
characterise the stochastic process, because in general the variables
$a(\rr,t)$ at different places/times will not be independent.  Thus
$P(a,\rr,t)$ is actually a marginal distribution of some more
complicated multivariate probability density.

A complete characterisation of the stochastic process would need an
infinite-dimensional probability distribution $P\bigl[a(\rr,t)\bigr]$
that gives the joint probability for all random variables $a(\rr,t)$.
However, one can usually avoid the difficulties associated with such a
distribution by considering the set of $n$-variable joint
distributions
\begin{equation}
  \label{eq:4}
  P_n(a_1,\rr_1,t_1,a_2,\rr_2,t_2,\ldots,a_n,\rr_n,t_n).
\end{equation}
Most stochastic processes \citep[\S 3.2.1]{Priestley1981} can be
completely specified by the set of all joint probabilities of the form
\eqref{eq:4} for all $n$ and all possible choices of
$\rr_1,t_1,\ldots,\rr_n,t_n$.  This should include all processes of
interest in physics and biology.  To define two-point correlation
functions we need only the set corresponding to $n=2$
\eqref{eq:two-point-dist}, which gives also the set for $n=1$ by
marginalising on the second variable.

The joint probability distributions that define the process are
time-translation invariant (TTI) if they obey (we omit the space
variables for clarity)
\begin{equation}
  P_n(A_1,t_1,\ldots,A_n,t_n) = P(A_1,t_1+s,\ldots,A_n,t_n+s)
\end{equation}
for all times, $s$ and $n$.  Stochastic processes obeying TTI are
called \emph{completely stationary processes.}  A related but less
restrictive notion is that of stationary processes up order $M$,
defined by the requirement that all the joint moments up to order $M$
exist and are time-translation invariant:
\begin{equation}
  \mean{ A^{m_1}(t_1)A^{m_2}(t_2)\ldots A^{m_n}(t_n) }= 
  \langle A^{m_1}(t_1+s)A^{m_2}(t_2+s)\ldots A^{m_n}(t_n+s)\rangle
\end{equation}
for all $s$, $n$, $\{t_1,\ldots,t_n\}$ and $\{m_1,\ldots,m_n\}$ such
that $m_1+m_2+\ldots+m_n\leq M$.  This is less restrictive not only
because of the bound on the number of random variables considered, but
also because invariance is imposed only on the moments, and not on the
joint distributions themselves.

\section{The estimator $\CC_k$ always has a zero}
\label{sec:zero-cross}

An important property of estimator \eqref{eq:conn-corr-estimate} will
always have zero, whatever the length of the sequence used to compute
it.  To see this, consider the quantity
\begin{equation}
B_i = (N-i) \hat C_{c,i} =  \sum_{j=1}^{N-i} \delta a_j \delta a_{j+i},
\end{equation}
and compute the sum
\begin{equation}
\sum_{i=0}^{N-1} B_i = \sum_{i=0}^{N-1} \sum_{j=1}^{N-i} \delta a_j
\delta a_{j+i} = \sum_{i=0}^{N-1} \sum_{j=1}^{N} \sum_{k=1}^{N} \delta a_j
\delta a_k \delta_{k,j+i}.
\end{equation}
But \(\sum_{i=0}^{N-1} \delta_{k,j+i}\) equals 1 if \(k\ge j\) and 0, so
\begin{multline}
\sum_{i=0}^{N-1} B_i = \sum_{j=1}^N\sum_{k=j}^N
\delta a_j \delta a_k = \frac{1}{2} \sum_{j\neq k}^N \delta a_j \delta
a_k + \sum_{j=1}^N (\delta a_j)^2 = \\
\frac{1}{2} \left[ \sum_{j=1}^N
\delta a_j \right]^2 + \frac{1}{2}\sum_{j=1}^N (\delta a_j)^2 =
\frac{1}{2}\sum_{j=1}^N (\delta a_j)^2 > 0,
\end{multline}
where the last equality follows because \(\sum_j \delta a_j=0\).
Now we can easily do the same sum starting from $i=1$:
\begin{equation}
  \sum_{i=1}^{N-1} B_i = -B_0 + \sum_{i=0}^{N-1} B_i =
  - \frac{1}{2}\sum_{j=1}^N (\delta a_j)^2 <0.
\end{equation}
This shows that at least some of the $B_i$ must be negative.  But
since $B_0>0$, the conclusion is that $B_k$, and hence $\hat C_{c,k}$,
which differs from it by a positive factor, must change sign at least
once for $k\ge1$.

\section{The shape of the correlation decay}
\label{sec:fits-time-corr}

\begin{figure}
  \centering
  \includegraphics{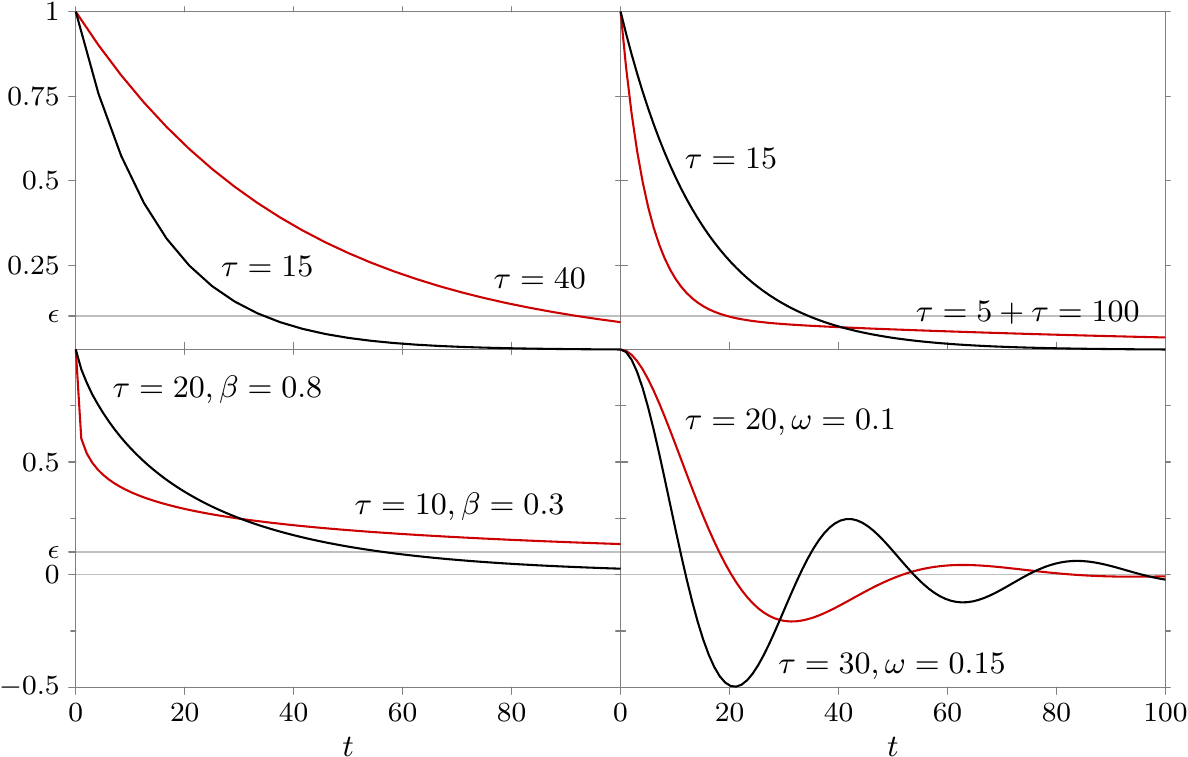}
  \caption{Different possible shapes for the decay of the time
    correlation. \textbf{Top left:} simple exponential.  \textbf{Top
      right:} simple exponential (black curve) and double exponential
    (red curve) decay.  In this case, the threshold criterion (here
    $\epsilon=0.1$) labels the red curve as the fastest, but it clearly
    has a longer tail.  \textbf{Bottom left:} stretched exponential.
    The integral correlation time \eqref{eq:tauint} is
    $\tauint \approx22.7$ (black curve), $\tauint\approx92.6$
    (red curve).  \textbf{Bottom right:} exponentially damped
    harmonic oscillations. }
  \label{fig:shapes}
\end{figure}

Fig.~\ref{fig:shapes} illustrates several (simple) possible shapes of
a decaying function.  It should make it clear why a threshold to
define a correlation scale is misleading unless the curves to be
compared are all the same shape.  Apart from the simple exponential
$\rho(t)=e^{t/\tau}$ and the sum of exponentials, the figure
illustrates two interesting cases.

One is the Kolrausch-William-Watts functions (lower left panel),
\def\rhokww{\rho_{\text{\tiny  K}}}
\def\taukww{\tau_{\text{\tiny  K}}}
\begin{equation}
  \rhokww(t) = e^{-(t/\taukww)^\beta},
\end{equation}
a function widely used to describe non-exponential decays, employing
only three parameters.  Here $\taukww$ is a time scale and $\beta$ is the
stretching exponent: the KWW function gives a stretched exponential
for $\beta<1$, and a compressed exponential for $\beta>1$.  Definition
\eqref{eq:tau-formaldef} gives 0 in the compressed case, and $\infty$
in the stretched case.  This can be understood considering the decay
as a superposition of exponential processes,
\begin{equation}
  \rhokww(t) = \int_0^\infty w(\tau) e^{-t/\tau} \, d\tau,
\end{equation}
which defines the correlation time distribution function $w(\tau)$.
It can be seen that the support of $w(\tau)$ extends to infinity for $\beta<1$
\cite{lindsey_detailed_1980}.  The integral correlation time is
instead more useful here,
\begin{equation}
  \tauint = \frac{\taukww}{\beta}\Gamma\left(\frac{1}{\beta}\right),
  \label{eq:avetau}
\end{equation}
where $\Gamma(x)$ is Euler's gamma function.  So $\tauint$ is
proportional to $\taukww$, but $\tauint$ incorporates information about
the shape of the tail of the decay, making it better suited to compare
two decays with different $\beta$.

The other case is that of damped oscillations, that appear in systems
with important inertial effects.  In the simplest harmonic form we have
\def\rhoosc{\rho_{\text{\tiny  osc}}}
\begin{equation}
  \rhoosc(t) = e^{t/\tau} \cos(\omega_0 t).
\end{equation}
The integral \eqref{eq:tauint} and spectrum \eqref{eq:HHrelaxtime}
scales are
\begin{align}
  \tauint &= \frac{\tau}{1+(\omega_0 \tau)^2},  &
  \tau_0 & = \frac{\tau}{\sqrt{1 + (\omega_0\tau)^2}}.
\end{align}
In the overdamped case, $\omega_0 \tau \ll 1$ both scales tend to
$\tau$, but in the opposite limit $\omega_0\tau\gg1$, where the
function oscillates many times before significant damping is observed,
$\tauint \sim 0$ while $\tau_0 \sim 1/\omega_0$.

\section{Algorithms to compute time correlations}
\label{sec:algor-comp-time}

The estimators can be computed numerically by straightforward
implementation of equations \eqref{eq:conn-corr-ns-est}
or~\eqref{eq:conn-corr-estimate}, although in the stationary case it
is much more efficient to compute the connected correlation through
relation \eqref{eq:power-spec} using a fast Fourier transform (FFT)
algorithm.  Let us focus on the stationary case and examine in some
detail these algorithms.

Algorithm \ref{algo:direct} presents the direct method.  It is
straightforward to translate the pseudo-code to an actual language of
your choice.  Apart from some missing variable declarations, the only
thing to consider is that it is probably inconvenient (or even
illegal, as in classic C) to return a large array, and it is better to
define $C$ as an output argument, using a pointer or reference (as
e.g.\ FORTRAN or C do by default) to avoid copying large blocks of
data.  The advantages of this algorithm are that it is self-contained
and simple to understand and implement.  Its main disadvantage is
that, due to the double loop of lines 8--11, it runs in a time that
grows as $N^2$.  For $N$ up to about $10^5$,
algorithm~\ref{algo:direct} is perfectly fine: a good implementation
in a compiled language should run in a few seconds in a modern
computer.  But this time grows quickly; in the author's computer
$N=5\cdot10^5$ takes 35 seconds, for $N=10^6$ the time is two and a
half minutes.  In contrast, the algorithm with FFT takes 1 second for
$N=10^6$ and 11 seconds for $N=10^7$.

\begin{algorithm}
  \caption{Compute the connected correlation of sequence $a$ (of
    length $N$) using the direct $O(N^2)$ method.  The connected
    correlation is returned in vector $C$.}
  \label{algo:direct}
  \begin{algorithmic}[1]
    \Function{timecorr}{$a$,$N$}
    \State $\mu \gets 0$   \Comment{Compute average}
    \For {$i=1,\ldots,N$} \State $\mu \gets \mu+a_i$ \EndFor
    \State $\mu \gets \mu/N$
    \For {$i=1,\ldots,N$} \Comment{Clear $C$ vector}
    \State $C_i \gets 0$ \EndFor

    \Statex

    \For {$i=1,\ldots,N$} \Comment{Correlation loop} \State
    $d \gets a_i-\mu$ \For {$k=0,\ldots,N-i$} \State
    $C_{k+1} \gets C_{k+1} + d * (a_{i+k}- \mu)$ \EndFor \EndFor

    \Statex
   
    \For {$i=1,\ldots,N$} \Comment{Normalize and return} \State
    $C_i \gets C_i / (N-i-1)$ \EndFor \State \textbf{return} C
    \EndFunction
  \end{algorithmic}
\end{algorithm}

If the correlation of really long sequences is needed, the FFT-based
algorithm, though more difficult to get running, pays off with huge
savings in CPU time at essentially the same numerical precision.  The
idea of the algorithm is to compute the Fourier transform of the
signal, use \eqref{eq:redu-almost-eq} to obtain the Fourier transform
of the connected correlation, then transform back to obtain $C_c(t)$.
This is faster than algorithm~\ref{algo:direct} because the clever FFT
algorithm can compute the Fourier transform in a time that is
$O(N\log N)$.

Actually, we need discrete versions of the Fourier transform formulas
(as we remarked before, the Fourier transform of the continuous time
signal does not exist).  The \emph{discrete Fourier transform (DFT)} and its
inverse operation are defined \cite[\S 12.1]{Press1992a} as (it is
convenient to let the subindex of $a_i$ run from $0$ to $N-1$ to write
the following two equations),
\begin{equation}
  \tilde a_k = \sum_{j=0}^{N-1} e^{2\pi i j k/N} a_j, \qquad
  a_j =\frac{1}{N} \sum_{k=0}^{N-1} e^{- 2\pi i j k/N} \tilde a_k,
\end{equation}
where we note that the inverse DFT effectively extends the sequence
periodically (with period $N$).  The discrete version of
\eqref{eq:redu-almost-eq} is \cite[\S 13.2]{Press1992a}
\begin{align}
  \tilde D_k & = | \tilde a_k |^2, & \text{where} &&
  D_j & = \sum_{k=0}^{N-1} a_k a_{k+j},
\end{align}
and where the definition of $D_j$ makes use of the (assumed)
periodicity of $a_i$.  $D_j$ is almost our estimate
\eqref{eq:conn-corr-estimate}: we only need to take care of the
normalization and of the fact that due to the assumed periodicity of
$a_i$ some past times are regarded as future, e.g.\ for $k=10$, in the
sum there appear the terms $a_0a_{10}$ up to $a_{N-11}a_{N-1}$ (which
are fine), but also $a_{N-10}a_0$ through $a_{N-1}a_9$, which we do
not want included.  This is fixed by padding the original signal with
$N$ zeros at the end, i.e.\ setting $a_k=0$ for $k=N,\ldots,2N-1$ and
ignoring the values of $D_j$ for $j\ge N$.

In summary, to compute the connected correlation using FFT the steps
are i) estimate the mean and substract from the $a_i$, ii) add $N$
zeros to the end of the sequence, iii) compute the DFT of the
sequence, iv) compute the squared modulus of the transform, iv)
compute the inverse DFT of the squared modulus, v) multiply by the
$1/(N-i)$ prefactor.  Pseudocode for this algorithm is presented as
algorithm~\ref{algo:FFT}.

\begin{algorithm}
  \caption{Compute the connected correlation of sequence $a$ (of
    length $N$) using a fast Fourier transform.  This algorithm is
     $O(N\log N)$.}
  \label{algo:FFT}
  \begin{algorithmic}[1]
    \Function{timecorr}{$a$,$N$} 
    \State $\mu \gets 0$ \Comment{Compute average}
    \For {$i=1,\ldots,N$}
    \State $\mu \gets \mu+a_i$ \EndFor
    \State  $\mu \gets \mu/N$
    \For {$i=0,\ldots,N$} \Comment{Substract the average from signal}
    \State $a_i \gets a_i-\mu$
    \EndFor
    \For {$i=N,\ldots,2N$}\Comment{Pad with 0s at the end}
    \State $\text{a}_i \gets 0 $
    \EndFor
    \Statex

    \State $b\gets$\Call{FFT}{$a$,$2N$} \Comment{Compute the FFT of a
      as a vector of length $2N$} \For {$i=1,\ldots,2N$}
    \Comment{Compute squared modulus of $b$} \State
    $b_i\gets \lvert b_i\rvert^2$ \Comment{Note that the Fourier
      transform is complex} \EndFor \State
    $C \gets$\Call{IFFT}{$b$,$2N$} \Comment{Inverse FFT} \Statex
 
    \State $C\gets$\Call{resize}{$C$,$N$} \Comment{Discard the last
      $N$ elements of $C$} \For {$i=1,\ldots,N$} \Comment{Normalize
      and return}
    \State $\text{C}_i \gets \text{C}_i / (N-i-1)$ \EndFor \State
    \textbf{return} C \EndFunction
  \end{algorithmic}
\end{algorithm}

To translate this into an actual programming language the comments
made for algorithm~\ref{algo:direct} apply, and in addition some extra
work is needed for lines 10--13.  First, one needs to choose an FFT
routine.  The reader curious about the FFT algorithm can read for
example \cite[Ch.~12]{Press1992a} or \cite{duhamel1990}, but writing
an FFT routine is not easy, and implementing a state-of-the-art FFT is
stuff for professionals.  Excellent free-software implementations of
the FFT can be found on Internet.  FFTW \cite{frigo2005}, at
\texttt{http://www.fftw.org} deserves mention as particularly
efficient, although it is a large library and a bit complex to use.
Note that some simpler implementations require that $N$ be a power of
two, failing or using a slow $O(N^2)$ algorithm if the requiriment is
not fulfilled.  Also pay attention to i) the difference between
``inverse'' and ``backward'' DFTs (the latter lacks the $1/N$ factor),
ii) how the routine expects the data to be placed in the input array,
iii) how it is returned, and iv) whether the transform is done ``in
place'' (i.e.\ overwriting the original data) or not.  If the routine
is a ``complex FFT'' it will expect complex input data (so that for
real sequences you will have to set to zero the imaginary part of the
$a_i$), while if it is a ``real FFT'' routine it will typically
arrange (``pack'') the output data in some way, making use of the
discrete equivalent of the $\tilde A(-\omega)=A^*(\omega)$ symmetry so
as to return $N$ real numbers instead of $2N$.  This affects the way
one must compute the squared modulus (lines 11--12).  For example, for
the packing used by the FFTW real routines, lines 11--12 translate to
(in C)
\begin{lstlisting}
  b[0]*=b[0];
  for (int i=1; i<N; i++) {
      b[i] = b[i]*b[i] + b[2*N-i]*b[2*N-i];
      b[2*N-i] = 0;
  }
  b[N]*=b[N];  
\end{lstlisting}

\bibliography{main}

\end{document}